\documentclass{article}

\usepackage{arxiv}

\usepackage[utf8]{inputenc} 
\usepackage[T1]{fontenc}    
\usepackage{hyperref}       
\usepackage{url}            
\usepackage{booktabs}       
\usepackage{amsfonts}       
\usepackage{nicefrac}       
\usepackage{microtype}      
\usepackage{lipsum}		
\usepackage{graphicx}
\usepackage{natbib}
\usepackage{doi}

\usepackage{amsmath,amsfonts,amsthm} 

\usepackage{subcaption}
\usepackage{nomencl}
\makenomenclature

\usepackage{tabularx} 
\usepackage{booktabs}
\usepackage{float} 
\usepackage{longtable}

\usepackage{url}              
\usepackage{color}            

\usepackage{authblk}

\title{Designing translational animal experiments by Bayesian meta-analytic predictive approaches}

\author{ \href{https://orcid.org/0000-0000-0000-0000}{\includegraphics[scale=0.06]{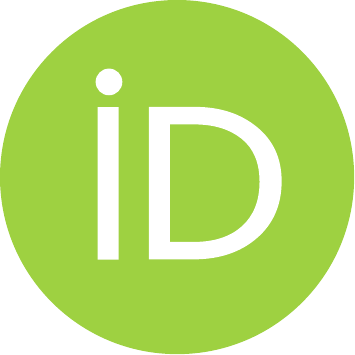}\hspace{1mm}Theresa Unseld}\thanks{theresa.unseld@uni-ulm.de} \\
	Department of Epidemiology and Medical Biometry\\
	Ulm University\\
	Ulm, Germany \\
}

\renewcommand{\shorttitle}{Designing translational animal experiments by Bayesian MAP approaches}

\hypersetup{
pdftitle={Designing translational animal experiments by Bayesian meta-analytic predictive approaches},
pdfauthor={Unseld ~Theresa, Elias D.~Striatum},
pdfkeywords={Translational research, Bayesian statistics, meta-analysis, design analysis, sample size calculation},
}

\begin{document}
\maketitle

\begin{abstract}
    The planning and conduct of animal experiments in the European Union is subject to strict legal conditions.
    Still, many preclinical animal experiments are only poorly designed. As a consequence, discoveries that are made in one animal experiment, cannot be reproduced in another animal experiment or discoveries in translational animal research fail to be translated to humans.    
    When designing new experiments in a classical frequentist framework, the sample size for the new experiment is chosen with the goal to achieve at least a certain statistical power, given a statistical test for a null hypothesis, a significance threshold and a minimally relevant effect size. The statistical test is a function of the data and the test is used to make statistical inference concerning the data's underlying, unobserved parameters of interest. In a Bayesian framework, inference is made by a combination of both the information from newly observed data and also by a prior distribution, that represents a priori information on the parameters. In translational animal experiments, a priori information is present in previously conducted experiments to the same outcome in similar animals.
    The prior information can be incorporated in a systematic way in the design and analysis of a new animal experiment by summarizing the historical data in a (Bayesian) meta-analysis model and using the meta-analysis model to make predictions for the data in the new experiment. This is called meta-analytic predictive (MAP) approach. In this work, concepts of how to design translational animal experiments by MAP approaches are introduced and compared to classical frequentist power-oriented sample size planning. Current chances and challenges, that exist in the practical application of these approaches in translational animal research, are discussed.
    Special emphasis is put on the construction of prior distributions and sample size calculation by design analysis. The considerations are motivated by a real world translational research example.
\end{abstract}


\section{Introduction}
	
	Translational research constitutes a key element to the development of new methods and therapies in human medicine. Situated in the late stage of preclinical research, its aim is to translate findings from basic laboratory preclinical research into the clinical application as potential treatments of human diseases.   
    In translational animal research biological pathways concerning clinically relevant phenotypes or pathologies are examined. These pathways are typically complex constructs whose mechanisms cannot be fully revealed in a single research experiment. Hence, the success of translational research fundamentally depends on the sensitive contemplation of new insights from an experiment in light of past insights and gained expertise knowledge.
    In Bayesian analysis, prior knowledge is formally incorporated as probability distribution into the analysis of newly observed data and updated to a posterior distribution that is then used for Bayesian inference. Several authors have already emphasized the value of Bayesian statistics in preclinical research (see for example \cite{spiegelhalter2004bayesian, walley2016using, kramer2017reducing, bonapersona2019repair, yang2019bayesian, gelman2019slamming, novick2021mean}). Nonetheless, practical applications of Bayesian methods for planning and analyzing preclinical animal experiments are rare (see \cite{walley2016using,kramer2017reducing, bonapersona2019repair}).
    A challenge in the application of Bayesian methods is the specification of a prior distribution. Translational animal experiments are typically characterized by the fact that the sample sizes in the experiment's groups are kept low for animal welfare purposes (see \cite{mayer2018essential}), like outlined in the 3R concept by \cite{russel959principles}. Especially in this situation, the choice of prior distribution can have a major impact on the posterior inference. Setting up a good prior distribution, that accurately reflects a priori information, can help to stabilize posterior inference, derive more precise estimates, reduce the impact of single extreme observations and to indicate if there might be something wrong or unexpected in the measurements of the new data that results in a wide posterior distribution.
    However, without a transparent justification for the choice of a prior and without a good understanding of the impact of the prior distribution on posterior inference, there is a risk that the final conclusion of a Bayesian analysis might be not sensitive (enough) to evidence in the newly observed data. On the other hand, if the prior distribution is chosen to have essentially no weight on posterior inference as compared to the data in the new experiment, then a major aspect of Bayesian inference and its associated benefits over frequentist inference are missing.

    One systematic way of setting prior distributions is to specify them
    based on a meta-analysis of relevant literature or databases as suggested by \cite{neuenschwander2010summarizing,pullenayegum2011informed,rhodes2015predictive,turner2015predictive,bartovs2021bayesian}.
    A meta-analysis is a popular tool for summarizing information from several statistical experiments in a common statistical model and quantifying the variability in different experiments. As a requirement, the experiments all have to address the same question. This assumption is often reasonable for control groups that stem from a series of experiments that address the same phenotypical outcome. The meta-analysis model can be used to estimate predictive distributions for a new experiment. These predictive distributions can be used to derive prior distributions for the analysis of the new experiment. This approach to a prior specification and Bayesian panning and analysis of the new experiment is termed meta-analytic predictive approach (MAP) \cite{neuenschwander2010summarizing}. These MAP priors fall into a class of data-based priors and are the focus in this article. Other methods for prior specification are summarized by \cite{mikkola2021} and are briefly discussed in the discussion.
    The MAP approach is illustrated for the mean in control groups of clinical studies by \cite{neuenschwander2010summarizing}. However, they don't use the historical data to derive prior distributions for other model parameters like the variance.
    Furthermore, animal experiments are characterized by their own challenges and methods suggested in human clinical trials cannot be transferred in a straight-forward manner to the application in animal experiments without considering possible adaptations (see \cite{walley2016using, kramer2017reducing}). 
    Firstly, due to the small sample sizes in the experiments' groups, the estimates obtained from animal experiments are usually characterized by large uncertainties.
    A further challenge is that, although many animal experiments are conducted, the results from the analysis are usually unorganized and restricted to limited access (see \cite{kramer2017reducing, bonapersona2019repair, novick2021mean}). It has been the idea to initiate search tools to perform systematic reviews of preclinical studies (see \cite{bahor2021development}) or launch common big databases to gather the information from more institutions (see \cite{keenan2009best, beckers2009towards, maddatu2012mouse, mcentyre2015biostudies, steger2020introducing, pognan2021etransafe}). Until such databases are fully developed, the remaining alternative option is to construct prior distributions form the little historical information that is available and be aware of the present uncertainties about the model parameters' true values. 
    The usage of Bayesian methods allows to fit meta-analysis models also in the challenging situation a of few, small previous experiments. Methods for Bayesian meta-analysis in the context of few, small studies have recently been proposed for the application in humans by \cite{friede2017metaFewSmall}. Furthermore, the MAP approach has been discussed in the context of animal experiments by \cite{walley2016using}. Still, there exist only few applications of how such Bayesian meta-analytic predictive methods are used to analyze real-world examples from preclinical translational research and even fewer examples of how to conduct sample size calculations for animal experiments in a Bayesian framework (see \cite{kramer2017reducing}). 
    
    The planning and conduct of animal experiments in the European Union (EU) is subject to strict legal conditions as outlined for example in the EU directive 2010/63/EU or for Germany in the legislation for animal welfare (Tierschutz-Versuchstierverordnung) (see \cite{richtlinieEU2010}). In particular, animal experiments are subject to an authorization by the respective competent authorities. Researchers can apply for this authorization by submitting an animal experiment proposal. If the animal experiments are set up to proof hypothesis, the proposals must be submitted together with a form or with a biometric review that provide a description of the analysis methods as well as a justification of the proposed sample size.
    Classically, sample size calculations are done in a frequentist framework by choosing the minimal number of animals that reaches a predefined power of a statistical test for a null hypothesis, given a data model, including its parameters. 
    However, the classical approach to null hypothesis significance testing (NHST) and power analysis has been criticized for several reasons (see \cite{gelman2014beyond, kruschke2015doing, schonbrodt2018bayes, stefan2019tutorial,gelman2020regression}). 
    Claiming statistical significance by a frequentist test for a certain hypothesis is equivalent to claiming statistical significance based on the corresponding p-value of the test or the confidence interval (see \cite{lehmann2006testing}). 
    These decision rules for a null hypothesis have been criticized for hypothesis testing, for example by \cite{wagenmakers2020principle,gelman2020regression, schad2022workflow}. One critique is connected to the \emph{principle of predictive irrelevance} (see \cite{wagenmakers2020principle}) and Bayes factors are proposed and as an alternative.
    \cite{gelman2014beyond, gelman2020regression} point out that, by choosing an experimental design with the goal of reaching a certain power of the statistical (while limiting the probability of type I errors of the test), other important goals of the experimental design are missed. More specifically, they argue that, by focusing on power and statistical significance, the reported estimates are systematically biased. This is because actually small effects are only reported for those data sets where the observed effect size happens to be large (by chance) and those cases are highly subjective to being overestimated and to being of the wrong sign.
    Sample size calculations in a classical framework are based on fixed estimates of an effect size and of the additional model parameters. One option is to derive these estimates from a researcher's previous experiment. This is problematic because the small sample sizes in animal experiments result in parameter estimates that are accompanied with large uncertainties (see \cite{mayer2013limitierte}) and these uncertainties are not well reflected by single point estimates.
    As a consequence, the sample size calculations in animal experiments are often only of limited use.
    As an alternative to power-oriented sample size calculation, design analysis using fake-data simulation has been suggested by \cite{kruschke2015doing,gelman2020regression}.
    
     In this paper the above aspects of Bayesian analysis, meta-analysis, prior specification using predictive approaches and design analysis using fake-data simulation are considered jointly in the context of sample size calculation for translational animal experiments. To the author's knowledge, there exists so far no work that illustrates sample size calculation for translational animal experiments using these approaches jointly. 
     After the introduction in this section, section two introduces sample size calculations in a classical frequentist framework and introduces fake-data design analysis as its alternative. Furthermore, a Bayesian MAP approach to designing and analysing new experiments based on historical data is explained. A distributional Bayesian model is used to deal with unequal group variances in the historical and new data.
     The considerations are applied to a real-world application example section three. In section four the main points are summarized and possible extensions are discussed.

	
\section{Methods}
\subsection{Statistical hypothesis tests and power-oriented sample size calculation}\label{sec:sample_size}
  For sample size calculations, assumptions have to be made concerning the distribution models for all experimental groups and assumptions concerning the experimental design that specifies which comparisons are made. Furthermore, estimable effects of interest, $\delta$, have to be defined that typically represent the differences in the means or effects of two or several groups. Additionally, for making binary decisions
  whether there is sufficient evidence in the data that the true effect of interest $\delta$ is different from a null effect $\delta_0$ or not, a decision function is required. The decision function in the classical classical frequentist framework is a statistical hypothesis test $\varphi$. Statistical hypothesis tests $\varphi$ compare test statistics $T$ to a critical value $c$ to make a decision whether or not to reject a null hypothesis 

\begin{align}
\label{eq:h02sided}
    H_0 :\{\delta=\delta_0\} &\text{vs. } H_1 :\{\delta\neq\delta_0\}
\end{align}

If $H_0$ is rejected, the alternative hypothesis $H_1$ is said to be accepted. 
 The tests statistic $T$ is constructed as a function of the data whose probability distribution under $H_0$ is known or can be approximated by a known distribution. 
A common situation is the comparison of two independent groups, an experimental (E) and control (C) group (e.g. knockout animals vs. wildtyp animals), under assumption of normally distributed data:

\begin{align}
	\label{eq:data_new}
	&y_{ik} = \theta_i + \epsilon_{ik}, & \epsilon_{ik} \sim \mathcal{N}(0, \sigma_i^2), \ i=C,E, \ k=1,\ldots,n_i.
\end{align} 

\subsubsection{Hypothesis tests in a frequentist analysis}
As frequentist hypothesis test on the mean difference $\delta=|\theta_E-\theta_C|$ the two-sample t-test can be used in case of equal variances $\sigma_E=\sigma_C$ (homoscedasticity) (see \cite{Lehmann2022}). Often, however, the residual variance is greater in the experimental than the control group due to varying treatment effects and the Welch test (Satterthwaite test) by \cite{welch1938significance, satterthwaite1941synthesis} is more appropriate since it does not make a homoscedasticity assumption:

\begin{align}
	\label{eq:welchTest}
	T_\text{Welch}=\frac{\overline{y}_1.-\overline{y}_2.}{\sqrt{\frac{\sigma_C^2}{n_1}+\frac{\sigma_E^2}{n_2}}}
\end{align}

Under $H_0$, the distribution of the test statistic $T_\text{Welch}$ is approximated by a t-distribution $t_{\Tilde{\nu}}$ with modified number of degrees of freedom $\Tilde{\nu}$ (for details see \cite{welch1938significance, welch1947generalization}). The variances $\sigma_E^2$ and $\sigma_C^2$ can be estimated as empirical variances from observed data. 
The corresponding two-sided hypothesis test is then

 \begin{align}
 \label{eq:welchHypothesistest}
     \varphi_\text{Welch}=1\{|T_\text{Welch}|>t_{\Tilde{\nu};1-\frac{\alpha}{2}}\}
 \end{align}

where $t_{\Tilde{\nu};1-\frac{\alpha}{2}}$ denotes the $1-\frac{\alpha}{2}$ quantile of the t-distribution with modified degrees of freedom.
 There are two standard types of errors that can be made by a hypothesis test:
		
		\begin{enumerate}
			\item Type I error: $\varphi$ rejects $H_0$ although it is true
			\item Type II error: $\varphi$ doesn't reject $H_0$ although it is not true
		\end{enumerate}
	
Both errors cannot be minimized simultaneously. Since the type I error is generally regarded as worse then the type II error, a classical hypothesis test $\varphi=\varphi_\alpha$ is set up to limit the probability of type I error by a significance level $\alpha$ and then finding the optimal test among all the tests at level $\alpha$ by minimizing the probability of type II errors (see \cite{lehmann2006testing}). 
This gives a test $\varphi_\alpha^\ast$ that has maximal power $P(\varphi=1|H_1)$ while controlling type I errors at level $\alpha$. 
The sample size is then classically calculated as minimal number for which the chosen hypothesis test $\varphi_\alpha^\ast$ detects a clinically relevant effect size $\delta^\ast$ at least with probability $1-\beta$. 	
More specifically, in applications with simple analytic distribution form of the test statistic, the sample size can be calculated by solving the power inequality
		
		\begin{align}
			P(\varphi_{\text{Welch},\alpha,n_E,n_C}=1| \delta \geq \delta^\ast) \geq 1-\beta.
			\label{eq:Schranke Power}
		\end{align} 



 For the Welsh test, the critical value $t_{\Tilde{\nu};1-\frac{\alpha}{2}}$ itself is a functions of the sample size through the (approximated) degrees of freedom. Hence, the inequality cannot be solved directly. Alternative approaches are to approximate the critical values by a normal distribution or to use an iterative approach. 
 

\subsubsection{Hypothesis tests in a Bayesian analysis}
``Statistically significant" in a frequentist context means that the test statistic $T$ is greater than the critical value $c$ (here: $c=t_{\nu;1-\frac{\alpha}{2}}$) or equivalently that the p-value of the test is smaller than $\alpha$ or the confidence interval at level $1-\alpha$ for the tested effect $\delta$ does not include the null value $\delta_0$ (see \cite{lehmann2006testing}). 
Similar decision rules or ``tests" can be established in a Bayesian framework. In a full Bayesian model, all parameters are modelled as probability distributions with their own prior distributions. The idea of Bayesian analysis is to update these prior distributions to posterior distributions by the information in new data using Bayes' rule.
Details on prior distributions are given in section \ref{sec:prior_distributions}.
Endowing all model parameters with probability distribution rather than fixing parameters to concrete values leads on the one hand to better representation of uncertainties than in the frequentist framework. On the other hand it often leads to complicated posterior density functions that require the evaluation of high-dimensional integrals and can no longer be expressed in analytic form. Numerical computation using Markov chain Monte Carlo (MCMC) methods is a common alternative. In MCMC methods the posterior of the parameter distribution is approximated by a series of MCMC draws.
For one-sided test problems, like the upper test problem $H_0: \{\delta\leq \delta_0\}$ vs. $H_1: \{\delta > \delta_0\}$, the estimated posterior probability $p(\delta|y)$ of $\delta$ being greater than a null value $\delta_0$, given the data $y=(y_C,y_E)$, or than a clinically relevant value $\delta^\ast$ can be computed. To transform this idea into a decision rule this posterior probability can be compared to a predefined critical value like suggested by \cite{weber2019applying}. 
This approach cannot be used for the two-sided problem \eqref{eq:h02sided}, when the tested effect $\delta$ is modeled by a continuous probability distribution, like it is the case when $\delta=|\theta_E-\theta_C|$ reflects the difference in two continuous means. In this case the posterior probability of a single point event $P(\{\delta=\delta_0\})$ is equal to zero. Alternatives are the consideration of two-sided Bayes factors or Bayesian credible intervals.

Credible intervals are Bayesian versions of frequentist confidence intervals with slightly different interpretation. By definition, a frequentist $1-\alpha$ confidence intervals for $\delta$ is an interval with random bounds that covers the unknown (but regarded as fixed) parameter $\delta$ with probability $1-\alpha$. Hence, if one would generate data from the corresponding probability model and calculate a confidence interval for each data set, one would expect that $(1-\alpha)\%$ of these intervals would cover (the fixed) $\delta$. The interpretation of a $1-\alpha$ Bayesian credible interval is instead that it is set up (with bounds regarded as fixed) so the probability that a random realization of $\delta$ falls within the credible interval is $1-\alpha$. Equivalently it can be stated that the Bayesian  $1-\alpha$ credible interval includes $(1-\alpha)\%$ of the posterior probability mass of $\delta$, given the data $y$ (see \cite{held2014applied}). Yet, this definition does not uniquely define the interval. There are two common types of Bayesian credible intervals: quantile intervals and highest density intervals. The bounds of a $1-\alpha$ quantile interval for a parameter are given by the $\frac{\alpha}{2}$ and $1-\frac{\alpha}{2}$ quantiles of its posterior distribution and is also called equal-tail interval. In praxis, the bounds of a quantile interval can be approximated by the quantiles of the MCMC draws. A highest density interval (HDI) is defined as a credible interval for which the posterior density of each point inside the interval is higher than the posterior density for an arbitrary point outside the interval. This is a desirable property as a summary of the distribution and is especially relevant for skewed distributions where, for the quantile interval, it is possible that parameter values inside the quantile interval are less probable than parameter values outside the interval. Moreover, the HDI is the smallest among all possible credible intervals which is a desirable property when it is used for posterior inference. On the other hand, an advantage of the quantile interval is, that it is easier to interpret for transformed  parameters than the HDI. This is because one can derive the quantile interval of the transformed parameter simply by back-transforming the intervals that where derived for the transformed parameter. In contrast, the HDI of the untransformed parameter cannot be simply derived by back-transforming the HDI of the transformed parameter.
A further difference is, that the quantile interval always includes the median of the posterior distribution, whereas the HDI always includes the mode(s) of the posterior distribution (\cite{kruschke2015doing}).
In symmetric distributions, the quantile interval and the HDI return similar results.
Moreover, under certain conditions Bayesian credible intervals also coincidence frequentist confidence intervals coincidence (see \cite{jaynes1976confidence}). 
With Bayesian credible intervals one can set up a decision rule by testing if $\delta_0$ falls withing the $1-\alpha$ posterior credible interval of $\delta$. A more meaningful approach may be to test whether the posterior interval excludes a region of practical equivalence (ROPE) which may be defined as all parameter values that are smaller (in absolute value) than the minimal clinically relevant effect size $\delta_\text{rel}$, as suggested by \cite{kruschke2015doing}). As a further alternative, Kruschke suggests to set as goal not to reach a certain power but rather a certain precision. An according decision rule can be implemented by deciding if the width of the credible intervals is smaller than a threshold value representing a target precision.

	 
 Classical decision rules for a null hypothesis have been criticized for null hypothesis testing, for example by \cite{wagenmakers2020principle,gelman2020regression, schad2022workflow}. One critique is connected to the \emph{principle of predictive irrelevance}
   stating that data that are predicted equally well by both a null model $\mathcal{M}_0$ (corresponding to $H_0$) and an alternative model $\mathcal{M}_1$ (corresponding to $H_1$), data which the authors in \cite{wagenmakers2020principle} call uninformative or \emph{irrelevant}, should not lead to favor one model above the other. However, this can happen in the above described scenario of null hypothesis significance testing (NHST) when intervals or p-values are estimated only under one of both models. Instead, to quantify the evidence for an effect that differs from the null hypothesis, Bayes factors are proposed.
Bayes factors compare the probability of the observed data under (at least) two models $\mathcal{M}_0$, $\mathcal{M}_1$:

\begin{align}
    \textbf{BF}_{10}=\frac{p(y|\mathcal{M}_1)}{p(y|\mathcal{M}_0)}
\end{align}

  This Bayes factor gives an impression of how much more likely the data was generated by the model under $H_1$ over the model under $H_0$. It is related to the odds of posterior model probabilities $\frac{p(\mathcal{M}_1|y)}{p(\mathcal{M}_0|y)}$ by being the factor by which the ratio of prior model odds $\frac{p(\mathcal{M}_1)}{p(\mathcal{M}_0)}$ changes after observing the data:
  \begin{align*}
   \frac{p(\mathcal{M}_1|y)}{p(\mathcal{M}_0|y)}=   \textbf{BF}_{10} \cdot \frac{p(\mathcal{M}_1)}{p(\mathcal{M}_0)}
  \end{align*}
  
  If the Bayes factor is greater than one this indicates that, after observing the data, the odds for the model under $H_1$ over $H_0$ have increased as compared to the a priori expectations.  \cite{schad2022workflow} show that, to make this an approach that is sensible to the observed evidence in the data, it is essential to chose appropriate prior distributions for the model parameters. Typical methods for estimating Bayes factors based on the prior distributions and the data are bridge sampling (\cite{bennett1976efficient}) or the Savage-Dickey method (\cite{dickey1970weighted}). 
To set up decision rules that are based on Bayes factor, a threshold has to be defined as to when the null hypothesis is rejected. 
 \cite{lee2014bayesian} provide a rough interpretation scheme for Bayes factors that is adjusted from \cite{jeffreys1998theory}. They declare moderate evidence for $H_1$ if the Bayes factor $BF_{10}$ is greater than $3$. This can be used to calculate the percentage with at least moderate evidence for $H_1$ as a binary decision function. This classification and decision rule present a convenient overview and method to get something like a power estimate from the Bayes factors, but should not be used as a strict rule (see \cite{schad2022workflow,schonbrodt2018bayes}). Instead, the whole distribution of the Bayes factors should be considered and ideally simulation based calibration and sensitivity analysis, as presented by \cite{schad2022workflow}, should be carried out to ensure a correct interpretation of the Bayes factors.
Alternatively, a heuristic decision rule can also be defined by making a decision in favor of $H_1$ if this is the model with the higher posterior model probability $p(\mathcal{M}_i|y)$, $i=0,1$ (see \cite{schad2022workflow}).
This decision rule however does not include the outcome of no evidence for either hypothesis and is only optimal if both errors of the corresponding decision rule (deciding for $H_1$ when the data in fact corresponds to model $\mathcal{M}_0$ and deciding for $H_0$ when the data in fact corresponds to model $\mathcal{M}_1$) are equally bad. This is typically not the case in clinical or preclinical research.
A more principle scheme for deriving decision rules is by the definition and optimization of utility functions.
 Utilities define the cost or value of decisions, conditionally on the null or alternative hypothesis being true and are necessary to judge the performance of a decision function. In the frequentist framework utilities are defined in terms of type I and type II error rates and decision functions are constructed by bounding type I errors and optimization with regard to type II errors. Differences in the Bayes factor oriented decision rules and the frequentist hypothesis tests are, for example, that the Bayes factors can distinguish between no evidence and evidence for $H_0$, whereas frequentist tests can not.

 

	\subsection{Simulation based design analysis} \label{sec:simulation_based_design_evaluation}

 \cite{gelman2014beyond, gelman2019slamming,gelman2020regression} propose design analysis using fake-data simulation as an alternative to classical power-oriented sample size calculations.
Using fake-data simulation, the effect of varying parameters on the predefined decision functions can be examined in combination with relevant candidates for the sample size. To determine if predefined statistical goals are met, models are fitted to the data, statistical analysis are performed and the previously defined decision functions are evaluated. The statistical goals are formalized in utility functions as introduced in the previous section. 
Here, utilities are calculated as type I and type II error rates or false discovery rates (FDR) and true discovery rates (TDR) by determining the percentages how often the data were simulated with a $\delta$ equal to the null effect but the decision functions decided against the null effect and how often the decision function decide against the null effect when the data were simulated with $\delta$ corresponding to increasing effect sizes. The goals are then to (find a sample size to) reach certain TDRs or a certain power while limiting the FDR or type I error rates by a certain $\alpha$. Additional model characteristics are examined as the \emph{type S (sign)} errors, as the probability that the estimate of the true effect has the wrong sign, given that is statistically significant and \emph{type M (magnitude)} errors, as the probability that the effect estimate is greater in absolute value than the absolute value of the true effect, given that it is statistically significant \cite{gelman2014beyond}.
Moreover, the the mean-squared-error (MSE) of the estimate of the true effect size is examined in simulated data under varying true effect sizes by \cite{gelman2019slamming}.
 Also the distribution of Bayes factors is visualized and it is examined how often the Bayes factors are falsely greater or smaller than one, if the lower bound of the $95\%$ interval of the posterior model probability for the model under $H_1$ exceeds the value of $50\%$ and what percentage of the Bayes factors lies within the categories defined by \cite{jeffreys1998theory}.
Given the utility function(s) and decision rules, a minimal sample size can be chosen that reaches one or several of these predefined goals.

 $R=10000$ simulated data sets are constructed in accordance with model \eqref{eq:data_new} as sum of a random control group, a treatment effect and group-specific residuals. In this work the data in the new experiment is generated in a frequentist framework with fixed values for $\delta$, $\theta_C$, $\psi$ and $\lambda$ and randomness in the simulated data comes only from the group-specific residuals.

\begin{equation}
	\begin{aligned}
		\label{eq:datsimbvtv}
		&y^\ast_{ik,r}=\theta^\ast_{C,r}+\delta^\ast_r 1(i=E)+\epsilon^\ast_{il,r} & \\
		&\epsilon^\ast_{ik,r}  \sim \mathcal{N}(0,\sigma^{\ast 2}_i), \ \sigma^\ast_i=\exp(\eta^\ast_{\sigma_{i,r}}), \text{ where } \eta^\ast_{\sigma_{i,r}}= \psi^\ast_{r}+ \lambda^\ast_r1(i=E)\ \\
	\end{aligned}
\end{equation}

for animal $k=1,\ldots,\tilde{n}_i$ in group $i=\text{C,E}$ of data set $r=1,\ldots,10000$.
Alternatively, the new data could also be generated in a fully Bayesian framework as discussed in the discussion.

For each simulated data set a Bayesian model is fit with the \texttt{brms} function of the \texttt{brms} package (\cite{brms2021}) in \cite{r2021}, using the default MCMC parameters. 
 Frequentist point estimates and confidence intervals for the population effects are estimated with the \texttt{lm} function in base \texttt{R}. The steps and decisions that are commonly made in a Bayesian framework are described as a Bayesian workflow by \cite{gelman2020bayesian,schad2021toward}. After fitting a Bayesian model the next step in a Bayesian workflow is to validate the computations. To decide whether the MCMC draws are likely to have converged against the target posterior distribution, characteristics of the MCMC chains are examined in convergence diagnostics. More specifically,
MCMC trace plots, auto correlation plots, the effective sample size and the $\hat{R}$ statistic are examined (for details see \cite{gelman2013bayesian}).
 Bayesian credible intervals are estimated as highest density intervals with the \texttt{bayestestR} package (\cite{bayestestR}) and quantile intervals computed as empirical quantiles of the posterior MCMC draws. Bayes factors are computed with the \texttt{hypothesis} function of the \texttt{brms} package.
 To account for heteroscedasticity, a distributional model is fit in \text{brms} and
in the frequentist model, heteroscedasticity is accommodated by the estimation of robust confidence intervals with the \texttt{sandwich} (\cite{sandwich2006}) package that estimates heteroscedasticity consistent (HC) variances. More specifically, a HC3 type estimator is chosen in the frequentist design that is also appropriate for smaller sample sizes (see \cite{long2000using}). Additionally, frequentist p-values in the Welsh test are calculated.

\subsection{Meta-analysis model for the historical data} \label{sec:metaanalysis}

The distributions in the sampling model \eqref{eq:datsimbvtv} as well as the prior distributions for a Bayesian analysis of the simulated (and new) data are based on a Bayesian meta-analysis model of relevant historic data.
The hope is that, by the usage of Bayesian estimation and historical evidence, a prior knowledge and uncertainties are better reflected and a sample size can be found that is more likely to actually achieve the predefined statistical goals than with classical frequentist methods. In the best case, and if the prior distributions reflect (major aspects of) the simulated data correctly, using the historical information as prior distribution in the Bayesian analysis of the new data can even reduce the number of animals that are needed to reach the predefined statistical goals.

\subsubsection{Normal-Normal hierarchical model}\label{sec:NNHM}

The historical data are modeled as data from $G$ different animal experiments with $n_g$ animals each, $g=1,\ldots,G$. As simplest and most commonly assumed case the data are assumed to be normally distributed as 

\begin{align}
	\label{eq:likelihoodMA}
	&y_{gk}=\theta_g+\epsilon_{gk},&\epsilon_{gk} \sim \mathcal{N}(0,\sigma_g^2) & \ \ k=1,\ldots,n_g, \ g=1,\ldots,G.
\end{align}

with residuals $\epsilon_{gk}$, $g=1,\ldots,G$, $k=1,\ldots,n_g$, and experiment specific means $\theta_g$, $g=1,\ldots,G$ estimated as arithmetic means $\overline{y}_g=\frac{1}{n_g}\sum_{k=1}^{g} y_{gk}$, $g=1,\ldots,G$.
Data, for which a normal assumption is inappropriate, such as skew and non-continuous data, can often be transformed to resemble a normal distribution and be handled by this model (see \cite{hedges1985statistical, hartung2001refined, higgins2011cochrane}).
As an assumption that allows the usage of the historical data for the analysis of the new experiment, the new data and the historical data are assumed to be \emph{exchangeable}. This means that there are assumed to be no systematic differences in the new and the historic data. This assumption is modeled by a random-effects meta-analysis for the historical data and the usage of this model's mean prior predictive distribution to construct a prior distribution for the parameters in the new experiment as outlined by \cite{neuenschwander2010summarizing}. 
Heterogeneity as variance of the historical experiments may occur due to different ages, animals strains, laboratory conditions or measuring instruments. Such heterogeneity is modeled by heterogeneity components $\gamma_g$ that represent deviation from a common mean $\mu$ in the experiment specific means $\theta_g=\mu+\gamma_g$, $g=1,\ldots,G$.
The parameters $\mu$ and $\tau$ in the distribution of the experiment specific parameters $\theta_g$ of interest are referred to as \emph{hyperparameters}. 

A Normal-Normal hierarchical model (NNHM) for the historical data is formulated as

\begin{equation}
	\label{eq:NNHM}
	\begin{aligned}
		&y_{gk}|\theta_g,\sigma_g \sim\mathcal{N}(\theta_g,\sigma_g^2) \\
		&\theta_g \sim \mathcal{N}(\mu,\tau^2). 
	\end{aligned}
\end{equation}

This model is termed \emph{hierarchical} since it includes connected sampling distributions on two levels and Normal-Normal hierarchical model since a normal assumption is assumed for both the residuals $\epsilon_{gk}$ on the level of the individual data and the heterogeneity components $\gamma_g$ on the level of the experiment specific means. 
In a frequentist framework this model with normally distributed means $\theta_g \sim \mathcal{N}(\mu,\tau^2)$ is also referred to as random effects model. Fitting a meta-analysis in a Bayesian instead of a frequentist framework has proven to be beneficial especially in the case of small, few studies by \cite{friede2017metaFewSmall, friede2017meta2studies}.
The estimation of the group-specific means in a hierarchical model with a common mean as pooled effect leads to so called \emph{shrinkage estimates} that are shrunken towards the common mean $\mu$ as compared to the estimation of independent means in a fixed effect model. The advantage of the shrinkage estimation is that single extreme values are relativized and the uncertainty in single groups can be reduced by borrowing information from other groups. The degree of shrinkage of one group $g\in\{1,\ldots,G\}$ depends on its sample size $n_g$ (or the associated standard error $s_g$) and the between-group heterogeneity $\tau^2$. Shrinkage is higher for smaller $n_g$ (bigger $s_g$) and smaller $\tau^2$ (for details see \cite{gelman2013bayesian, schmidli2014robust,neuenschwander2016robust, wandel2017using, rover2021bounds}).

\subsection{Prior distributions}\label{sec:prior_distributions}

In a Bayesian framework further probability distribution models as prior distributions are set up for the additional parameters ($\sigma_g$, $\mu$ and $\tau$ in the meta-analysis in equation \eqref{eq:NNHM} model and $\theta_C$, $\delta$, $\psi$ and $\lambda$ in model \eqref{eq:datsimbvtv}).
A popular choice for a parameter's prior distribution is a so called \emph{conjugate prior} from the distribution family that is conjugate to the family of the modeled likelihood distribution of the data \cite{gelman2006prior}. Choosing such a conjugate prior distribution for a model parameter implies that also the parameter's posterior distribution is from the same family as the prior distribution. This has interpretational and computational advantages as outlined in \cite{gelman2006prior,gelman2013bayesian}.
A prior distribution can be categorized according to its information content to be either non-informative, weakly informative or informative. This categorization can be controlled by the prior distribution's variance parameter and has to be interpreted in context of the likelihood distribution the data (for details see \cite{gelman2013bayesian}).
Non-informative priors have minimal impact on the posterior distribution as compared to the impact of the data. They are constructed so that their probability density is flat relative to the probability density of the data. In contrast, informative prior distributions are designed to represent the full available a priori knowledge as accurately as possible and to have a major impact on the parameters' posterior distribution together with the impact of the data. Weakly-informative prior distributions constitute a compromise between non-informative and informative distributions. They are intentionally designed to be flatter than informative prior distributions. Weakly-informative priors can be chosen to have a regularizing functionality on the posterior distribution by restricting the posterior distribution to a plausible parameter range (for details see for example \cite{gelman2013bayesian, mcelreath2018statistical}).
 Generally, context specific weakly informative or informative priors are preferred over non-informative priors, especially when Bayes factors are estimated (see \cite{gelman2006prior,betancourt2017shape, seaman2012hidden, betancourt2017shape, schad2021toward, lemoine2019moving}).

\subsubsection{Variance parameters}
With few studies, special care hast to paid to the specification of a prior distribution for the heterogeneity parameter $\tau$ in model \eqref{eq:NNHMbvtv} (see \cite{gelman2006prior,friede2017metaFewSmall}). 
 \cite{friede2017metaFewSmall} recommend a prior distribution that puts most of their probability mass to areas that represent small to large heterogeneity and leave only a little fraction to values that represent a larger heterogeneity. The interpretation of the heterogeneity degree depends on the scale of the modelled parameter. \cite{spiegelhalter2004bayesian} suggest to classify heterogeneity in context with the residual standard deviation $\sigma$ of the data model into the following classes:

 \begin{table}[ht!]
	\centering	
	\begin{tabular}{lc} 
		\toprule
		Heterogeneity & $r:=\frac{\tau}{\sigma}$\\  
		\midrule 
		small & 0.0625  \\ 
		moderate & 0.125 \\
		substantial & 0.25  \\
		large & 0.5 \\
		very large & 1.0 \\
		\bottomrule
	\end{tabular}
	\caption{Classification of heterogeneity in relation to the standard deviation of the data according to \cite{spiegelhalter2004bayesian}.}
	\label{tab:Heterogenitaetsklassen}
\end{table}

Recommended prior distributions are then from the family of of folded non-central t-distributions with special cases of the half-t and half-Normal distribution. The half-Normal distribution $\mathcal{HN}(\varphi)$ has a scale parameter $\varphi$ and is related to a standard normal distribution with mean zero and variance $\varphi^2$ by taking the standard normal distribution's absolute values. Compared to the half-t distribution the tail of a half-normal distribution is smaller which puts less weight on extreme heterogeneity values (see \cite{spiegelhalter2004bayesian}). The distributions parameters in this applications example are set up to reflect the assumptions on the heterogeneity as classified by table \ref{tab:Heterogenitaetsklassen}. 


\subsubsection{Intercept parameters}
As prior distribution for the intercept parameters normal distributions are chosen which is, conditionally on all other model parameters, the conjugate family to the normal distribution of the data. More specifically, as starting point for the intercept's prior, a \emph{unit information prior (UIP)} is recommended (see \cite{kass1995reference, rover2021weakly}). 
A unit information prior has the information content (variance) that corresponds to one single typical data point. The unit information prior for the historical data is set up to be centered at the mean of the historical data. Strictly speaking orienting the prior distribution on information from the data implies to use the same information twice, once for setting up the prior and once through the likelihood of the observed data that both go into the estimation of the posterior distribution. This contradicts the Bayesian concept of a prior distribution as representation of \emph{a priori} knowledge before seeing the analyzed (historic) data. It can nonetheless serve as a starting point to ensure that the prior distributions are centered at some reasonable area. By choosing a variance parameter that leads to a wide enough but not unrealistically wide prior distribution, it reflects only a rough guess and lets the data contribute more exact information to refine the parameter's posterior distribution while having a regularizing functionality \cite{mcelreath2018statistical}. This again can be examine in sensitivity analysis.
A unit information prior can also be used as reference prior for testing Bayesian hypothesis or for an assessment of the effects on the posterior compared to more informative priors (see \cite{kass1995reference,raftery1998bayes,neuenschwander2020use,li2021rbestNormal, rover2021weakly}).

For the mean $\theta_C$ and the standard deviation $\sigma_C$ in the control group in the new experiment, MAP priors are computed. 
 Therefore the \texttt{posterior\_epred} function of the \texttt{brms} package is used to estimate the expected posterior predictive distribution in the operation groups. The expected posterior predictive distribution of the group without operation is used to derive parametric prior distributions for the mean and standard deviation $\theta_C$ and $\sigma_C=\exp(\psi)$ in the new experiment, as explained in the next section. The priors for $\theta_E$ and $\sigma_E=\exp(\psi+\lambda)$ are set up to be centered around the same values like the priors $\theta_C$ and $\sigma_C=\exp(\psi)$, but with higher variance parameters, corresponding to unit information priors. The higher variance parameters reduce the weight of the prior distributions as compared to the data. The intention for centering the prior distributions of the parameters in the control and experimental group around the same value is that, a priori, the two groups are expected to be equal on average. Meanwhile, the intention for the higher variance in the priors of the parameters in the experimental group is, that for the control group there is historical data available, so the prior distribution should have larger influence than for the experimental group, where no (directly related) historical data is available.

 \subsubsection{Approximation of the MCMC draws by parametric distributions}\label{sec:approximiation_of_the_MCMC_draws}

To incorporate the historical data as proper prior distributions, the non-parametric estimates of the posterior predictive distribution, as represented by the MCMC draws, are approximated by parametric distributions. \cite{rover2021weakly, rover2022summarizing} illustrate three general methods for fitting parametric distributions to MCMC draws. In each method the first step is to specify a distribution family. 
Choosing a (to the data distribution) conditionally conjugate normal prior distribution for the mean parameter in the historical data model ensures that the posterior distribution is from the same parameter family, conditionally on fixed values of all other model parameters. However, when the other model parameters are not fixed, the posterior distribution is not necessarily a simple normal distribution. In case of the NNHM from equation \eqref{eq:NNHM}, \cite{rover2017bayesian} show that, with a normal prior distribution for $\mu$ and a half-normal prior distribution for $\tau$, the posterior predictive distribution for the man parameter $\mu$ and the parameter $\theta^\ast$ in the new experiment are normal-mixture distributions. Thus, also more complex distribution families have to be considered for the approximation of the posterior MCMC draws. After a distribution family is specified, a simple method to estimate its parameters is by taking point estimates, like the mean of median, from the posterior draws. This approximation of the posterior draws as point estimates however is a reduction of information and does not always reflect all important characteristics of the posterior distribution. An alternative method is to approximate the posterior draws by marginal distributions (see \cite{rover2021weakly}). A third alternative is to choose a model family and determine its parameters by maximum likelihood (ML) estimation, the expectation-maximization (EM) algorithm or moment matching (MM). Different candidates for distribution families can be compared by using estimators of the model fit or of their predictive performance like the Akaike information criterion (AIC) \cite{weber2019applying}. In the Bayesian context, the Watanabe-Akaike information criterion (WAIC) and Leave-one-out cross-validation (LOO-CV) are preferred over AIC, as outlined by \cite{gelman2014understanding,vehtari2017practical}. But since the parametric distributions are fit with frequentist and not with Bayesian methods, WAIC and LOO-CV are not considered here.
In this work, the parametric distribution candidates for the MAP priors are normal, normal mixture and t-distributions. For fitting a mixture distribution, the \texttt{automixfit} function of the \texttt{RBesT} R package (\cite{weber2019applying}) is used. This function uses the EM algorithm to fit a series of normal mixture distributions with increasing number of mixture components and selects the best fit according to the AIC value (which penalizes model complexity). For comparison, simple normal and non-centered t-distributions are fit using ML estimation. Prior predictive checks, as described in section \ref{sec:prior_distributions}, are perfomred to ensure the priors are reasonable.

The approach to formulate a predictive distribution as prior distribution in the new experiment is termed meta-analytic predictive (MAP) approach (\cite{neuenschwander2010summarizing}) and requires the explicit formulation of a prior distribution as predictive distribution, given the historical data (MAP prior). An alternative to this sequential approach for the analysis of the historical and new data is the meta-analytic-combined (MAC) approach where both historic an new data are analyzed in a common analysis. \cite{schmidli2014robust} show that the MAP approach is theoretically equivalent to the MAC approach.
In practice, the MAC approach has the advantage to be more direct and easier since it requires only to fit one single Bayesian model instead of one model for the historical data, one for the derivation of a MAP prior from the historic data and one for the analysis of the new data. In the context of design analysis and sample size determination, however, the MAP approach has the advantage of allowing better judgment and control over the influence of the historic data on the estimation results in the new analysis. Furthermore, an \emph{effective sample size (ESS) of the MAP prior} $n_\text{eff}$ can be calculated that quantifies the influence and information content of the historical data as ratio of the variance of the MAP prior in a heterogeneous sample to the variance of the MAP prior in a homogeneous, pooled sample. $n_\text{eff}$ gives an estimate of the number of animals that can be saved in the new experiment by using the information in the historical data (given that the new and historical data are exchangeable) (see \cite{neuenschwander2010summarizing, neuenschwander2020predictively}).

\section{Application example}
As an application the simulation based design analysis and sample size determination is carried out for an animal experiment from translational preclinical research that aims to examine the role of the C5aR1 receptor on bone quality under postmenopausal osteoporosis in mice. The pathological condition of postmenopausal osteoporosis is realized by ovariectomy in mice. Bone quality is measured (among other indicators) by the (unit-less) relative bone volume (bone volume to tissue volume (BV/TV)) in a $\mu$CT scan. This experiment is referred to as \emph{C5aR1 experiment} in the following. 
The aim of the planned animal experiment is to test whether or not there is evidence for an association between the C5aR1 knockout and the relative bone volume in mice. If there exists such an associations, then a C5aR1 knockout may serve as basis for a potential treatment that can be examined in humans in clinical trials.
The planned experiment shall consist of data from twelve week old female mice of the C57BL/6J strain, which is the standard mouse strain from the Jackson laboratory research institution. 

\subsection{Data and models in the application example}

\subsubsection{Historical data}
Internal historical data from a previous experiment is available from the proposer for planning the new experiment. This internal data comprises data from twelve ovariectomized mice and ten mice with Sham operation. Yet, the internal historic data from the proposer represents only a fraction of the information that has been collected so far regarding bone quality in mice since bone quality is an active research topic in preclinical research (see for example \cite{ignatius2011complement, ignatius2011anaphylatoxin,modinger2018c5ar1} and the citations therein).
The inclusion of additional data has the potential of providing more information about the distribution of the relative bone volume in mice, the variability among and between different experimental groups and upon which effect sizes are realistic. However most of the available literature cannot be used as further historical data, since the animals characteristics (such as age and the animal strain) of the external data and the experimental conditions under which external data have been collected, deviate from the internal historic data or the outcomes are measured with different methods, have different definitions or scales.
\begin{sloppy}
   An easy and comprehensive option for retrieving control data is the Mouse Phenotype Database (MPD) (\cite{grubb2004collaborative,bogue2018mouse}). It is an open-access database that collects phenotype data for the characterization of inbred mice. The data can be used for characterizing the correlations of complex traits and as control data or for characterization of mutation effects (see \cite{consortium2007Integration}). The MPD data comes in a tidy, standardized format and also the experiment protocols and tools for data analysis are provided. The data is made available by worldwide researchers and managed by employees of the MPD, who also endow the data with a common public ontology like the Mammalian Phenotype Ontology, that was introduced by \cite{smith2005mammalian}, what makes it easier identify and compare relevant data for the outcome of interest. 
\end{sloppy}	

The search term ``BV/TV" on the MPD web-page leads data from 31 strains of Collaborative Cross (CC) wildtyp mice from an experiment of \cite{levy2015collaborative}. CC mice are recombinant inbred mice from eight genetically divergent strains. They are characterized by a high degree of genetic diversity that represent on average $90\%$ of the allelic diversity in the whole mouse genome \cite{chesler2008collaborative}. In the planned experiment the mice shall undergo ovariectomy. Since ovariectomy is expected to have an impact on the relative bone volume, the MPD mice cannot be used on its own to estimate a posterior predictive distribution for the mean relative bone volume in the new experiment. Still, the estimation of the distribution for the mean in the wildtyp mice can give an idea for the range of plausible effect sizes that are examined in the design analysis. For example, one might expect that the best one can expect from the C5aR1 knockout in the experimental group in the new experiment is to completely reverse the negative effect of the ovariectomy and bring the relative bone volume back to the basic level in wildtyp mice.
Moreover, the additional consideration of the external data can help to quantify heterogeneity in the outcome in different strains and ideally, it could also give an impression of how representative the internal mice strains are for the mouse genome in general with respect to the outcome relative bone volume by comparing internal wildtyp mice to the MPD wildtyp mice. However, int his case there is no internal data from wildtyp mice but only ovariectomized and Sham mice. 

The external data and internal data are represented in table \ref{tab:bvtvMAPData}. Since the hypothesis refers only to female animals, all male animals from the external MPD data set are excluded from the meta-analysis.
The relative bone volume can by definition only take positive values. In the historical data most of the values were close to zero with a couple of very big values. A logarithmic transformation was applied to the data to transform the scale of the outcome to the real numbers and to make it resemble more a normal distribution.


\begin{table}[ht]
    \centering
    \begin{tabular}{lllrrrrr}
        \hline
        Experiment & strain & OP & n & $\widehat{\text{E(BV/TV)}}$ & $\widehat{\text{SD(EI)}}$ & $\widehat{\text{E}(\log(\text{EI}))}$ & $\widehat{\text{SD}(\log(\text{EI}))}$ \\ 
        \hline
        Internal & C57BL/6 & Ovx & 12 & 1.5 & 1.10 & 0.1 & 0.93 \\ 
        Internal & C57BL/6 & Sham & 10 & 4.2 & 1.19 & 1.4 & 0.36 \\ 
        MPD & PreCC1061/Tau & None & 2 & 9.2 & 0.74 & 2.2 & 0.08 \\ 
        MPD & PreCC111/Tau & None & 9 & 14.4 & 3.23 & 2.6 & 0.22 \\ 
        MPD & PreCC1156/Tau & None & 2 & 15.8 & 0.21 & 2.8 & 0.01 \\ 
        MPD & PreCC1513/Tau & None & 7 & 29.2 & 6.47 & 3.4 & 0.21 \\ 
        MPD & PreCC188/Tau & None & 9 & 6.9 & 2.11 & 1.9 & 0.29 \\ 
        MPD & PreCC1912/Tau & None & 6 & 12.8 & 1.17 & 2.5 & 0.10 \\ 
        MPD & PreCC2126/Tau & None & 6 & 2.4 & 2.14 & 0.6 & 0.76 \\ 
        MPD & PreCC2156/Tau & None & 8 & 7.7 & 2.74 & 2.0 & 0.34 \\ 
        MPD & PreCC2391/Tau & None & 2 & 2.9 & 1.76 & 0.9 & 0.66 \\ 
        MPD & PreCC2573/Tau & None & 3 & 12.5 & 9.42 & 2.4 & 0.70 \\ 
        MPD & PreCC2680/Tau & None & 7 & 5.8 & 2.29 & 1.6 & 0.53 \\ 
        MPD & PreCC2689/Tau & None & 6 & 6.5 & 2.24 & 1.8 & 0.33 \\ 
        MPD & PreCC2750/Tau & None & 5 & 6.6 & 3.91 & 1.8 & 0.61 \\ 
        MPD & PreCC3348/Tau & None & 7 & 15.1 & 3.91 & 2.7 & 0.24 \\ 
        MPD & PreCC3438/Tau & None & 5 & 7.2 & 3.46 & 1.9 & 0.49 \\ 
        MPD & PreCC3480/Tau & None & 3 & 6.8 & 7.06 & 1.4 & 1.40 \\ 
        MPD & PreCC3912/Tau & None & 10 & 11.3 & 2.32 & 2.4 & 0.20 \\ 
        MPD & PreCC4052/Tau & None & 12 & 13.5 & 11.47 & 2.3 & 0.81 \\ 
        MPD & PreCC4141/Tau & None & 6 & 12.8 & 5.47 & 2.5 & 0.46 \\ 
        MPD & PreCC4438/Tau & None & 3 & 8.5 & 0.67 & 2.1 & 0.08 \\ 
        MPD & PreCC4457/Tau & None & 7 & 14.9 & 3.83 & 2.7 & 0.30 \\ 
        MPD & PreCC519/Tau & None & 6 & 15.6 & 7.55 & 2.6 & 0.48 \\ 
        MPD & PreCC521/Tau & None & 7 & 6.1 & 1.47 & 1.8 & 0.23 \\ 
        MPD & PreCC557/Tau & None & 4 & 5.3 & 1.72 & 1.6 & 0.35 \\ 
        MPD & PreCC611/Tau & None & 3 & 8.7 & 0.16 & 2.2 & 0.02 \\ 
        MPD & PreCC670/Tau & None & 3 & 14.9 & 3.53 & 2.7 & 0.23 \\ 
        MPD & PreCC711/Tau & None & 3 & 12.3 & 1.01 & 2.5 & 0.08 \\ 
        MPD & PreCC72/Tau & None & 6 & 8.1 & 1.88 & 2.1 & 0.23 \\ 
        \hline
    \end{tabular}
    \caption{Sample size ($n$), mean ($\hat{\text{E}}$) and empirical standard deviation ($\hat{\text{SD}}$) of the relative bone volume (BV/TV) in the historical data on the original and the logarithmic scale. The estimates are calculated by the group factors experiment (internal, MPD), mouse strain and operation group (OP as ovariectomy (Ovx), Sham operation and no operation).}
    \label{tab:bvtvMAPData}
\end{table}

\subsubsection{Meta-analysis model}
A hierarchical or meta-analysis model with individual data (IP-MA, individual patient meta-analysis) (see for example \cite{lyman2005strengths,michiels2005random,van2010individual}) is fit in a Bayesian framework by using the \texttt{brm} function in the \texttt{brms} package (\cite{brms2021}).
For comparison, a frequentist model is fit in R with the \texttt{lme} of the \texttt{nlme} package (\cite{nlme2022}) to accommodate the random strain effects, where restricted maximum-likelihood (REML) estimates are derived by maximization of the log-restricted maximum-likelihood method (see \cite{pinheiro2006mixed}).
An operation group variable \texttt{OP} is included as predictor to indicate if the mice had underdone ovariectomy, a Sham operation or no operation. The ovariectomy and Sham operation may not have the same impact on all animals in the respective operation group whereas the data in the animals without an operation is expected to have smaller variance. To allow the mice in the ovariectomy group to have a different residual variance than the residual variance in the Sham operation and allow the residual variance to be yet different than the residual variance in the group without operation, a distributional model (like explained in \cite{burkner2020estimating}) is fit, in which the residual variance is modeled by its own operation group specific predictor term, similar to the one in equation \eqref{eq:datsimbvtv}.
Ideally, since bone quality is known to depend on age, this variable should be included as predictor term. But since each operation group is from a different age interval (the youngest are the animals without operation, and the animals with Sham operation and ovariectomy are all of an older age), the effects of age and the operation group cannot be untangled by the inclusion of the age variable.
The same problematic applies to the experiment where the data came from (internal, MPD). Also this variable should ideally be modeled as fixed or random effects parameter. But since both Sham and ovariectomized animals are internal data and the animals without operation are external data from the MPD also this effect cannot be estimated.

In summary, the the following meta-analysis model is fit:

	\begin{equation}
		\label{eq:NNHMbvtv}
		\begin{aligned}
			&y_{ijk} =  \alpha+
			\beta_1 1(i=1)+ \beta_2 1(i=2)+ \nu_j+\epsilon_{ijk},\\
			&\text{with operation group specific residuals}\\
			& \epsilon_{ijk}\sim \mathcal{N}(0,\sigma_i),\
			\sigma_i=\exp(\eta_{\sigma_i}), \\
			&\eta_{\sigma_i}=\psi +\lambda_1 1(i=1)+
			\lambda_2 1(i=2)\\
			&\text{and additional independent variance components}\\
			&\nu_j \sim  \mathcal{N}(0,\tau^2_\nu)\\
		\end{aligned}
	\end{equation}

where $k=1,\ldots,n_{ij}$ indices the mouse in operation group  $i=0,1,2$ (none, ovariectomy (Ovx),Sham) and strain $j=1,\ldots,22$. The choice of the prior distributions is explained in the next section.

\subsubsection{Prior distributions for the historical data}

In the \texttt{brms} package non-informative or weakly informative prior distributions are specified as default. Different default priors are chosen for the intercept (intercept of the mean, $\alpha$, and intercept of the logarithmic residual standard deviation, $\psi$), than for the \emph{population parameters}  that apply to the whole data (\emph{fixed effects} in a frequentist framework, here $\beta_i$, $\lambda_i$, $i=1,2$ and $\gamma$), and for group-specific \emph{variance parameters} that model the heterogeneity between groups (\emph{random effects} in a frequentist framework, here $\nu_j$, $j=1,\ldots,22$).

In table \ref{tab:priors_bvtvhistorical} the manually chosen prior distributions for parameters from model \eqref{eq:NNHMbvtv} are summarized and contrasted with the default priors in the \texttt{brm} function.
For the intercept, an unit information prior (UIP) is considered as reference as explained in section \ref{sec:prior_distributions}. The intercept $\alpha$ represents the mean relative bone volume in the animals without operation and at age of eight weeks on the logarithmic scale. An UIP is set up with information content corresponding to a single observation. Therefore the historical animals without operation are considered. Using ML estimations to fit a normal distribution to the group without operation, the residual standard deviation is estimated to be $0.72$. To make this a bit less informative, the standard deviation for the prior of $\alpha$ is increased to the value $1$. The mean of $\alpha$'s prior, is set according to the estimated mean in the group without operation which is $2$. These choices result in a $95\%$ interval of $[0.1,4]$ on the logarithmic scale and $[1.1,53]$ on the original scale.  
The operation-group specific residual standard deviations $\sigma_i$, $i=0,1,2$ (no operation, Ovx, Sham), in model \eqref{eq:NNHM} are modelled as exponent of a normally distributed linear predictor $\eta_{\sigma_i}$, that is centered at mean $0$. With a scale parameter of $0.5$, $\sigma_0$ has a $95\%$ quantile of $2.3$, which is considered as more realistic than the default $t_3(0,2.5)$ distribution in \texttt{brms} that leads to a $95\%$ quantile of $360$ of $\sigma_0$.

For the heterogeneity parameter $\tau_\nu$ of the variance parameters $\nu_j$, $j=1,\ldots,22$ (representing the variance due to the different strains) a half-normal prior is chosen as discussed in section \ref{sec:prior_distributions}. The scale parameter of $\tau_\nu$ is set to $\frac{1}{2}$, which corresponds to large heterogeneity when classified with table \ref{tab:Heterogenitaetsklassen} and the mean of the prior $\sigma_0$ as reference scale. A priori, large heterogeneity is expected to exist in the mice strains since they include CC mice from genetically diverse backgrounds as explained above.
The population effects $\beta_i$, and $\lambda_i$, $i=1,2$ are kept at their default values in \texttt{brms} and only modified if there are indications of convergence problems or if the prior predictive checks indicate that they are unrealistic. Neither is the case here.

\begin{table}[ht]
	\centering
	\begin{tabular}{lll}
		\hline
		Parameter & Default & Manual \\ 
		\hline & & \\[-1.5ex] 
		$\alpha$ & $t_3(1.7,2.5)$ & $\mathcal{N}(2,1^2)$ \\ 
		$\psi$ & $t_3(0,2.5)$ & $\mathcal{N}(0,0.5^2)$ \\ 
		$\beta_i$ & $\mathcal{U}(-\infty,\infty)$ & $\mathcal{U}(-\infty,\infty)$ \\ 
		$\lambda_i$ & $\mathcal{U}(-\infty,\infty)$ & $\mathcal{U}(-\infty,\infty)$ \\ 
		$\tau_\nu$ & $\mathcal{U}(-\infty,\infty)$ & $\mathcal{HN}(0,0.5^2)$ \\ 
		\hline
	\end{tabular}
	\caption{Prior distributions (default in \texttt{brms} and manual choices) for the Bayesian meta-analysis model of the historical data on logarithmic scale.} 
	\label{tab:priors_bvtvhistorical}
\end{table}

\subsubsection{Prior predictive checks}
 The model including the candidates for the prior distributions are tested in prior predictive checks, introduced by \cite{good1950probability}. In prior predictive checks a large amount of replication data is simulated from the given model and prior distributions with the aim to decide if the model and parameter distributions are (biologically) plausible. Although plausibility statements can already be made by reviewing the model and parameter distribution definitions, the interplay of all model components is most easily viewed and judged in such prior predictive checks where the generated data reflects all these model components and can be compared to a researchers prior expectation of how typical data in this context should look like. 
 To perform the prior predictive checks data sets are generated from model \eqref{eq:NNHMbvtv} with prior distributions as specified in \ref{tab:priors_bvtvhistorical}. 
 For the population effects $\beta_i$, $\gamma$, $\lambda_i$, $i=1,2$, that are necessary to construct the data in the ovariectomy and Sham operation groups, improper, non-informative priors $\mathcal{U(-\infty,\infty)}$ were chosen. Since no sampling is possible from this distribution and to reduce the scope of generated plots, the prior predictive checks are only presented for the group without operation. Hypothetical data for the other groups could be generated with priors that are approximating the $\mathcal{U(-\infty,\infty)}$ distribution, for example with a normal distribution $\mathcal{N}(0,\sigma^2)$ with very large $\sigma$.
 $1000$ prior predictive data sets are generated. For each data set the model parameters are simulated with random number generating functions in \texttt{R} and then the hypothetical data are constructed by the model described in \eqref{eq:NNHMbvtv}. For the animals in the group without operation this corresponds to

 \begin{align}	\tilde{y}_{r1jk}=\tilde{\alpha}_r+\tilde{\nu}_{r,j[l]}+\tilde{\epsilon}_{r1jk}
\end{align}

for mouse $k=1,\ldots,K$ from mouse strain $j[k]=1,\ldots,22$ in the hypothetical data set $r=1,\ldots,1000$. The number of mice per simulated data set, $K$, is chosen to correspond to the number of mice in the historical data which is $K=72$. However, the aim is not to choose prior distributions that exactly reflect the distribution of the historical data but rather to select weakly-informative prior distributions that lead to a prior predictive distribution that is less informative (flatter) compared to the actual observed historical data distribution but that excludes values that seem implausibly high in context of the historical data.
The theory is that, for a big number of generated data sets, the empirical distribution of the simulated data approximates the prior predictive distribution of the data (see \cite{good1950probability}). 


\subsection{Results from the application example}

\subsubsection{Fitting the Bayesian Normal-normal hierarchical model to the historic data}
\begin{figure}[ht]
    \centering
    \includegraphics[ height=0.6\textheight]{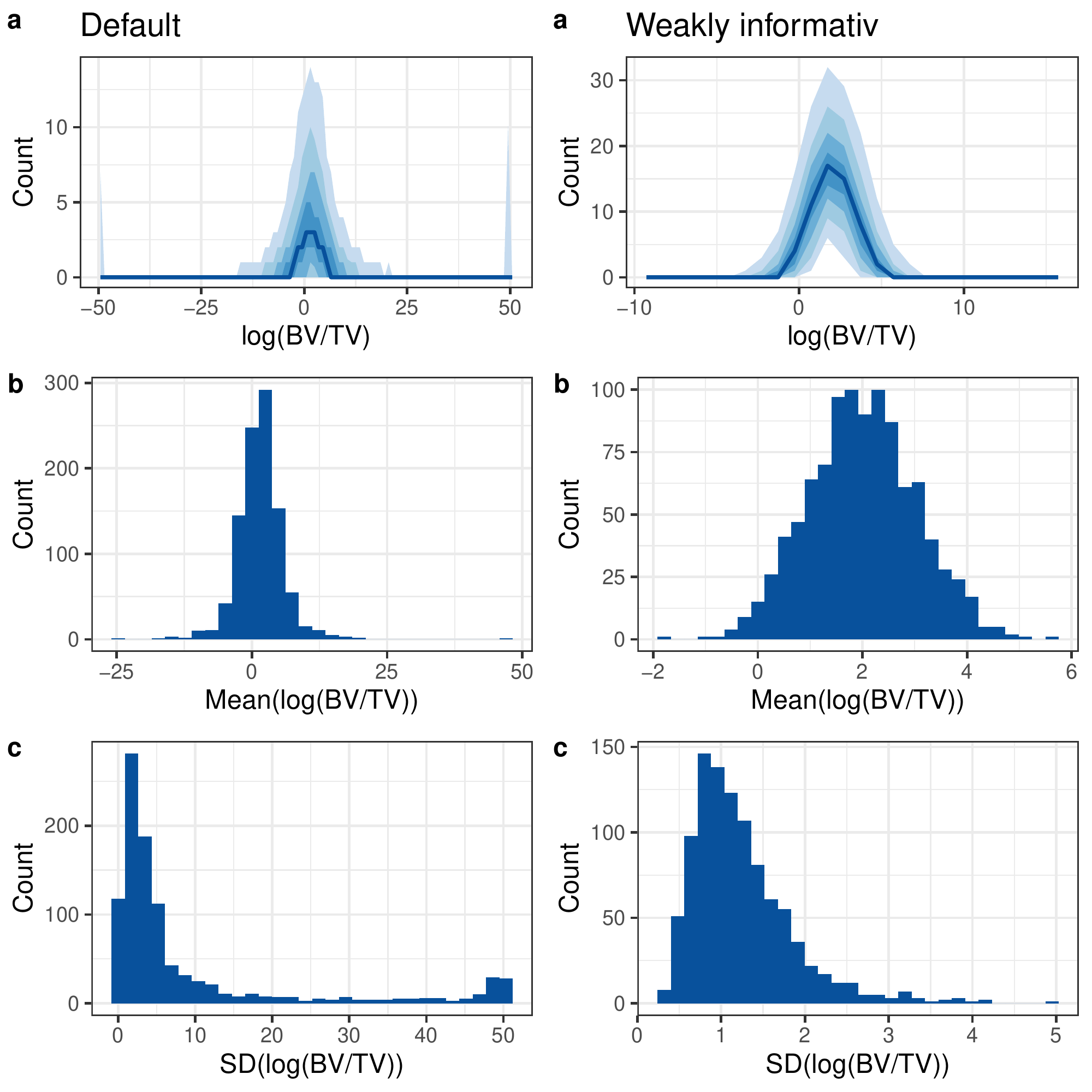}
    \caption{Graphical prior predictive checks adapted from \cite{schad2021toward} for the relative bone volume in animals without operation on a logarithmic scale with different prior distributions. Left column: Default prior distribution in the R package \texttt{brms}. Right column: weakly-informative prior distribution from table \ref{tab:priors_bvtvhistorical}. The predictive distributions were calculated over 1000 simulated data sets. a) Distribution of histograms calculated per simulated data set. The colored areas correspond in the order of increasing intensity to 10-90, 20-80, 30-70 and 40-60 percent intervals over all histogram frequencies of the simulated data sets. The dark curve in the middle of the intervals represents the distribution of the median over all simulate data sets. b) Distribution of arithmetic means. c) Distribution of standard deviations. Extreme log(BV/TV) values $<-50$ or $>50$ are represented as $-50$ and $50$ for representation.}
    \label{fig:pPcheck_fitbrms}
\end{figure}

The results of the prior predictive checks for the prior distributions of the meta-analysis model in the historical data are presented in figure \ref{fig:pPcheck_fitbrms}. The distribution of the histograms indicates that the default prior distributions in \texttt{brms} lead to very big values of the logarithmic relative bone volume in the group without ovariectomy. The $10\%$ and $90\%$ interval boundaries are almost at $-25$ and $25$, what corresponds to values about zero and $7.2 \cdot 10^{10}$ on the original scale. With the weakly-informative prior instead the $10\%$ and $90\%$ interval boundaries are at $-7.5$ and $7.5$, corresponding still to quite low and high values $5.5\cdot 10^{-4}$ and $1808$ on the original scale. From a biological perspective, the weakly-informative prior distributions seem more reasonable but still allow the actually observed data to have a major impact on the posterior distributions. Since no convergence problems in the model fit occur and since the posterior estimates seem reasonable and since the meta-analysis model is not the main model but rather is intended mainly for building a prior itself for the analysis of the new experiment, no further fine tuning of the prior distributions for the meta-analysis model is performed. However, additional prior distributions that result in overall smaller relative bone volumes could be examined. The plausibility of the estimated meta-analysis model is examined in posterior predictive checks as presented below. Furthermore, MCMC diagnostics are examined. The diagnostics showed no signs of divergence or high auto-correlations.

\begin{figure}[ht]
    \centering
    \includegraphics[ height=0.8\textheight]{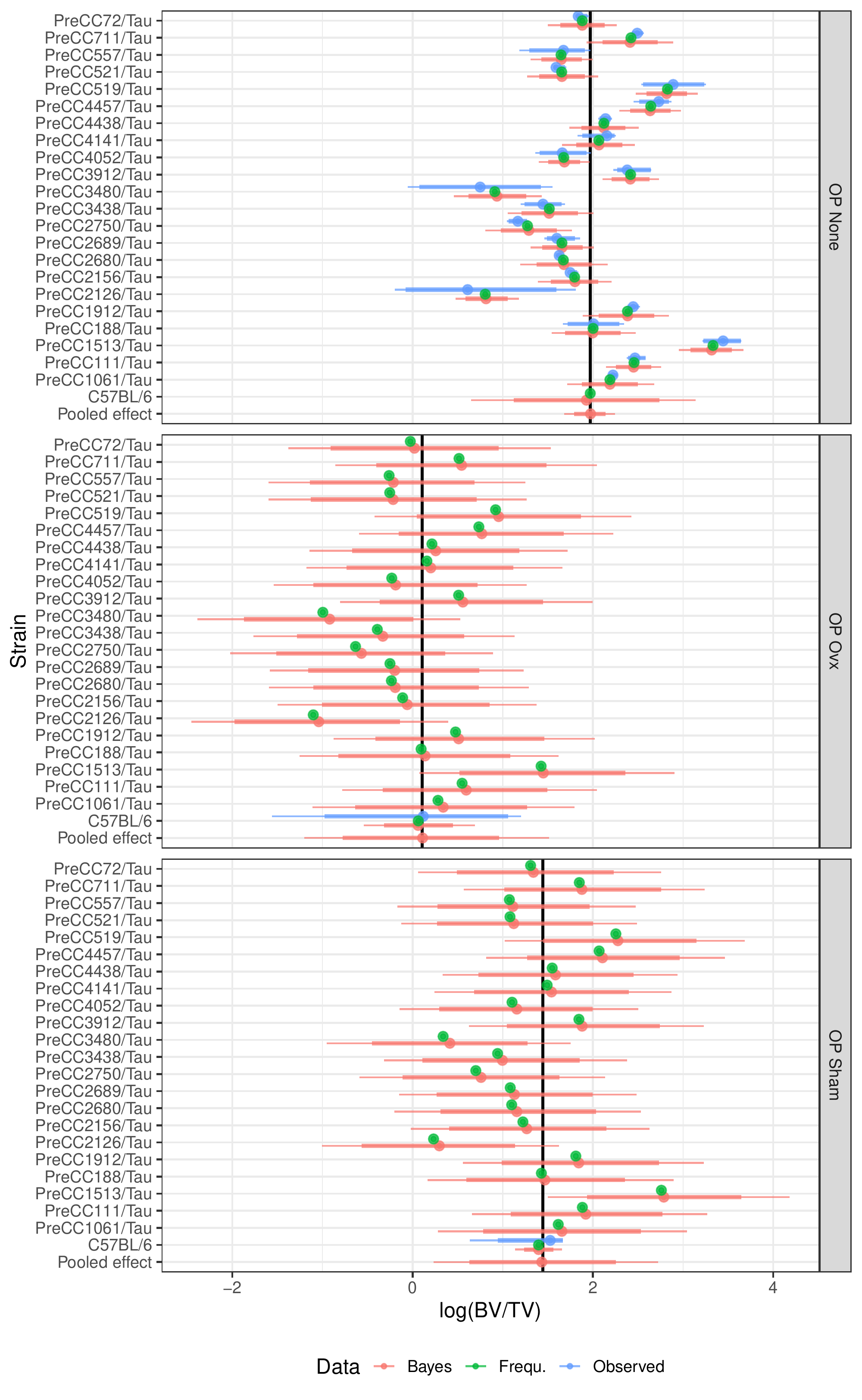}
    \caption{Forest plot with strain specific means of the historical animals and their common mean (pooled effect) stratified by the operation group (none, ovariectomy (Ovx), Sham). Bayes: Bayesian estimates as means of the posterior MCMC draws (points). Frequ.: frequentist REML estimates. Observed: strain-specific means in the observed data. The intervals are presented as $80\%$ (thick lines) and $85\%$ (thin lines) quantile intervals.}
    \label{fig:forestplot}
\end{figure}

The estimated population effects (fixed effects) and standard deviations of the variance components (random effects) and residuals in the meta-analysis model of the historical data are presented in table \ref{tab:bvtv_sds} as means with $95\%$ quantile intervals and compared with the estimates from a frequentist analysis with REML estimators and approximate confidence intervals under normal assumption. 
Overall, the estimations from the Bayesian MCMC and frequentist REML method are similar. Slightly bigger differences exist in the estimation of the residual standard deviations.
Notably larger differences exist for the residuals standard deviations of the Sham and Ovx group that have only very few observations. 
A forest plot of the strain effects is represented in figure \ref{fig:forestplot}. The hierarchical (random effects) model leads to group-specific estimates that are shrunken towards the common mean (pooled effect) and have smaller variance than in an independent estimation as fixed effects model, as discussed in section \ref{sec:NNHM}. The point estimates of the Bayesian and frequentist analysis are again quite similar.



\begin{table}[ht]
    \centering
    \begin{tabular}{lrl}
        \hline
        Variable & Bayes & Frequ. \\ 
      \hline
    Intercept & 2 [1.7,2.2] & 2 [1.7,2.3] \\ 
      Ovx & -1.9 [-3.2,-0.46] & -1.9 [-3.3,-0.51] \\ 
      Sham & -0.53 [-1.8,0.78] & -0.57 [-1.9,0.74] \\ 
      Strain & 0.64 [0.47,0.87] & 0.63 [0.45,0.88] \\ 
      SD(Sham) & 0.41 [0.254,0.689] & 0.36 \\ 
      SD(Ovx) & 1 [0.666,1.62] & 0.93 \\ 
      SD(None) & 0.37 [0.301,0.451] & 0.35 \\ 
       \hline
    \end{tabular}
    \caption{Estimated population effects (fixed effects) and standard deviations of the variance components (random effects) and group-specific residuals. Bayes: arithmetic means with $95\%$ quantile intervals of the MCMC simulations. Frequ.: frequentist REML estimators where the confidence intervals of the random effects are approximate confidence intervals under normal assumption.} 
    \label{tab:bvtv_sds}
\end{table}

\begin{figure}[ht]
	\centering
	\includegraphics[ height=0.6\textheight]{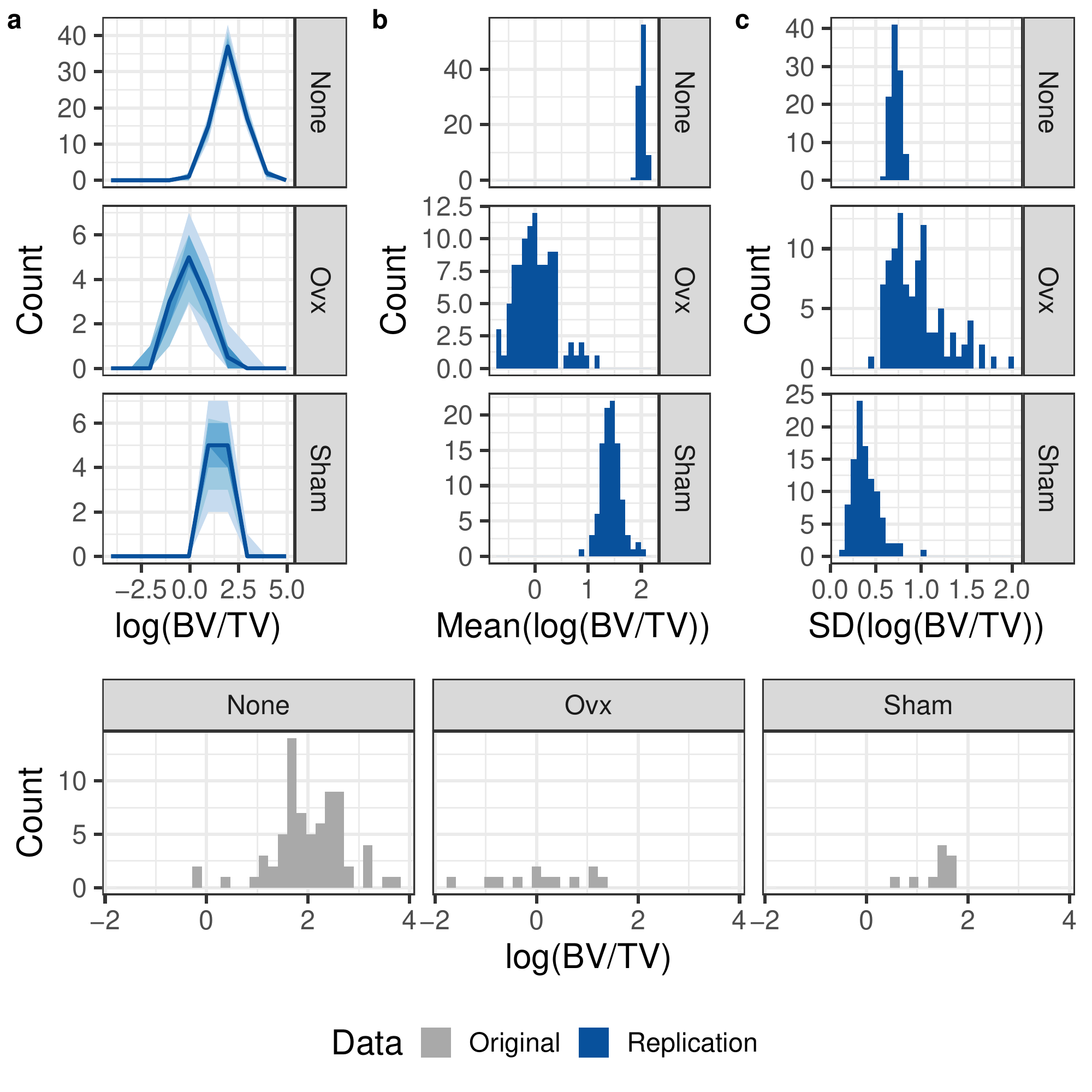}
	\caption{Graphical posterior predictive checks adapted from \cite{schad2021toward} for the relative bone volume in animals without operation on a logarithmic scale. Grey: original data. Blue: replicates from the MCMC fit in 1000 simulated data sets. a) Distribution of histograms calculated per simulated data set. The colored areas correspond in the order of increasing intensity to 10-90, 20-80, 30-70 and 40-60 percent intervals over all histogram frequencies of the simulated data sets. The dark curve in the middle of the intervals represents the distribution of the median over all simulate data sets. b) Distribution of arithmetic means. c) Distribution of standard deviations. Extreme log(BV/TV) values $<-50$ or $>50$ are represented as $-50$ and $50$ for representation.}
	\label{fig:ppchecks}
\end{figure}

As part of a Bayesian workflow, posterior predictive checks are performed to ensure that the model generates reasonable data in light of the original data (for details see \cite{gelman2013bayesian}). The results are presented in figure \ref{fig:ppchecks}. The posterior distributions seam reasonable in light of the observed data. Compared to the respective prior distributions (here only shown for the group without operation in figure \ref{fig:ppchecks}), the posterior distributions have smaller tails due to the updated information by the historic data. 

 \subsubsection{Approximation of the MCMC draws and definition of prior predictive distributions}

The results of the approximations of normal distributions by the ML method and of (according to AIC best) normal mixture distributions by the EM method and selection by the AIC (as desribed in section \ref{sec:approximiation_of_the_MCMC_draws}) are presented in table \ref{tab:fit_pars_muCT}. The approximations with both methods look quite similar and hence the approximation by a simple normal distribution with the ML method is selected as prior in place of a more complicated mixture distribution to avoid overfitting and for simplification, since it requires less parameter than the mixture distribution with more than one component.

\begin{table}[ht]
	\centering
	\begin{tabular}{llllrrrl}
		\hline
		Variable & Distribution & Method & Component & Weight & E & SD & Median (95\% interval) \\ 
		\hline
$\mu_C$ & Normal-Mix & EM & 1 & 1.00 & 0.10 & 0.69 & 0.1 [-1.2,1.5] \\ 
  $\mu_C$ & Normal & ML & 1 & 1.00 & 0.10 & 0.69 & 0.1 [-1.3,1.5] \\ 
  $\sigma_C$ & Normal-Mix & EM & 1 & 0.53 & 1.10 & 0.27 & 1.1 [0.71,1.7] \\ 
  $\sigma_C$ & Normal-Mix & EM & 2 & 0.47 & 0.90 & 0.15 & 0.9 [0.65,1.2] \\ 
  $\sigma_C$ & Normal & ML & 1 & 1.00 & 1.00 & 0.24 & 1 [0.64,1.6] \\ 
  $-\beta_1$ & Normal-Mix & EM & 1 & 1.00 & 1.90 & 0.70 & 1.9 [0.49,3.2] \\ 
  $-\beta_1$ & Normal & ML & 1 & 1.00 & 1.90 & 0.70 & 1.9 [0.53,3.3] \\ 
		\hline
	\end{tabular}
	\caption{Results from the approximation of the MCMC posterior distribution in the ovariectomized animals by parametric distributions. EM: (according to the AIC best) fit normal-mixture approximation of a series of models fitted by the expectation-maximization algorithm. ML: normal distribution fit by the maximum-likelihood method. E and SD: mean and standard deviation of the respective parametric distribution. 95\% interval: 0.025 and 0.975 quantiles of the parametric distribution.} 
	\label{tab:fit_pars_muCT}
\end{table}

An effective sample size of the normally distributed prior $p(\theta_C)$ for the control (Ovx) group is calculated with the \texttt{ess} function of the \texttt{RBesT} package by \cite{weber2019applying}. The calculation requires the specification a \emph{reference scale} as an estimate of the (within-group) residual standard deviation in the historic and new control group animals. This residual standard deviation estimate is set to the mean of the estimated posterior distribution of the residual standard deviation in the historical ovariectomized animals. The resulting estimate of the effective sample size of the prior is quite low with $n_\textbf{eff}=2$ indicating that, in this example, the benefit in using the information of the historical data in the analysis of the new experiment is only small. More informative prior distributions and models for the design analysis could be derived if there was more historical data available. Methods to promote the availability of historical data are described in the discussion.


\subsubsection{Design analysis and sample size determination}
 The candidates for the true $\delta$ in the simulated new data are taken from a range that spans from the minimum zero (corresponding to no effect, i.e. the mean in the control and experimental group are equal on average) to a maximum that corresponds to the mean of the parametric distribution that was fit to the negative difference in predicted means of the ovariectomized animals and the animals without operation ($\beta_1$ in table \ref{tab:fit_pars_muCT}). The mean of the parametric distribution was estimated to be $1.9$ and its standard deviation to be $0.7$. With respect to the estimated mean standard deviation in the ovariectomized group $\hat{\sigma}_C=\exp(\hat{\log(\sigma_C)})=1$, this estimate corresponds to a Cohen effect by \cite{cohen1988statistical} size of $d=\frac{1.9}{1}=1.9$ (very large to huge according to the classification heuristics by \cite{ferguson2016effect} and \cite{sawilowsky2009new}). 
 As further options, mean effect sizes of $\frac{1}{3}\cdot 1.9 \approx 0.6$ and $\frac{2}{3}\cdot 1.9 \approx 1.3$ are modelled that correspond to medium and large effect sizes with respect to $\hat{\sigma}_C=1$. Additional designs with treatment effect $\delta=0$ (no effect) are evaluated for investigating type I errors and false discovery rates. Furthermore, heteroscedastic designs with larger residual standard deviations in the experimental group than the control group (that might occur due to varying effects of the treatment) are examined. Therefore, the coefficient $\lambda$ is set to $\log(1.5)$ to simulate a standard deviation in the experimental group that is $1.5$ times the standard deviation in the control group ($\sigma_E=\exp(\psi+\lambda)=1.5\sigma_C$). As sample size candidates typical sample sizes from translational animal experiments are chosen as five and ten animals in either control our experimental group. If in the experimental designs five animals in either group seem to be to few for achieving a certain statistic goal and ten animals seem too much, a more finely-tuned set of candidate sample sizes in an in-between range of five and ten can be examined. The designs are summarized in table \ref{tab:Designs_muCT}. 
 The parameters for the prior and data distribution of the mean $\theta_C$ and the residual standard deviation $\sigma_C$ in the new experiment's control group are set according to the estimated parameters from the ML fit of the normal distributions in table \ref{tab:fit_pars_muCT}. The prior for $\theta_E$ is chosen to be weakly informative unit information prior (UIP) with a mean that equals the mean of $\theta_C$ and a standard deviation that corresponds to the estimated mean of the standard deviation in the historical ovariectomized animals. Also the mean of the prior for $\eta_{\sigma_E}=\log(\sigma_E)$ is set equal to the mean of $\eta_{\sigma_C}$. The standard deviation of the prior for $\eta_{\sigma_E}$ is set higher than that of $\eta_{\sigma_C}$, to the value one. With these parameters, the prior for $\sigma_C$ is centered around the same value as the prior for $\sigma_E$, but has larger tails that allows also for more extreme values.
 In prior predictive checks these priors seem reasonable and weakly-informative enough to not overrule the new data. 
 

\begin{table}[ht]
\centering
\begin{tabular}{rrrrrrr}
  \hline
Model & $n_E$ & $n_C$ & $\mu$ & $\delta$ & $\log(\sigma_C)$ & $\frac{\sigma_E}{\sigma_C}$ \\ 
  \hline
1 & 5 & 5 & 0.1 & 0.0 & 0.0 & 1.0 \\ 
  2 & 10 & 5 & 0.1 & 0.0 & 0.0 & 1.0 \\ 
  3 & 10 & 10 & 0.1 & 0.0 & 0.0 & 1.0 \\ 
  4 & 5 & 5 & 0.1 & 0.6 & 0.0 & 1.0 \\ 
  5 & 10 & 5 & 0.1 & 0.6 & 0.0 & 1.0 \\ 
  6 & 10 & 10 & 0.1 & 0.6 & 0.0 & 1.0 \\ 
  7 & 5 & 5 & 0.1 & 1.3 & 0.0 & 1.0 \\ 
  8 & 10 & 5 & 0.1 & 1.3 & 0.0 & 1.0 \\ 
  9 & 10 & 10 & 0.1 & 1.3 & 0.0 & 1.0 \\ 
  10 & 5 & 5 & 0.1 & 1.9 & 0.0 & 1.0 \\ 
  11 & 10 & 5 & 0.1 & 1.9 & 0.0 & 1.0 \\ 
  12 & 10 & 10 & 0.1 & 1.9 & 0.0 & 1.0 \\ 
  13 & 5 & 5 & 0.1 & 0.0 & 0.0 & 1.5 \\ 
  14 & 10 & 5 & 0.1 & 0.0 & 0.0 & 1.5 \\ 
  15 & 10 & 10 & 0.1 & 0.0 & 0.0 & 1.5 \\ 
  16 & 5 & 5 & 0.1 & 0.6 & 0.0 & 1.5 \\ 
  17 & 10 & 5 & 0.1 & 0.6 & 0.0 & 1.5 \\ 
  18 & 10 & 10 & 0.1 & 0.6 & 0.0 & 1.5 \\ 
  19 & 5 & 5 & 0.1 & 1.3 & 0.0 & 1.5 \\ 
  20 & 10 & 5 & 0.1 & 1.3 & 0.0 & 1.5 \\ 
  21 & 10 & 10 & 0.1 & 1.3 & 0.0 & 1.5 \\ 
  22 & 5 & 5 & 0.1 & 1.9 & 0.0 & 1.5 \\ 
  23 & 10 & 5 & 0.1 & 1.9 & 0.0 & 1.5 \\ 
  24 & 10 & 10 & 0.1 & 1.9 & 0.0 & 1.5 \\ 
   \hline
   \end{tabular}
	\caption{Designs (parameter for the simulated data and sample size candidates) for the design analysis for the outcome relative bone volume on a logarithmic scale.} 
	\label{tab:Designs_muCT}
\end{table}

\begin{figure}[ht]
	\centering
	
	\begin{subfigure}[b]{0.55\textwidth}
		\includegraphics[ width=1\linewidth]{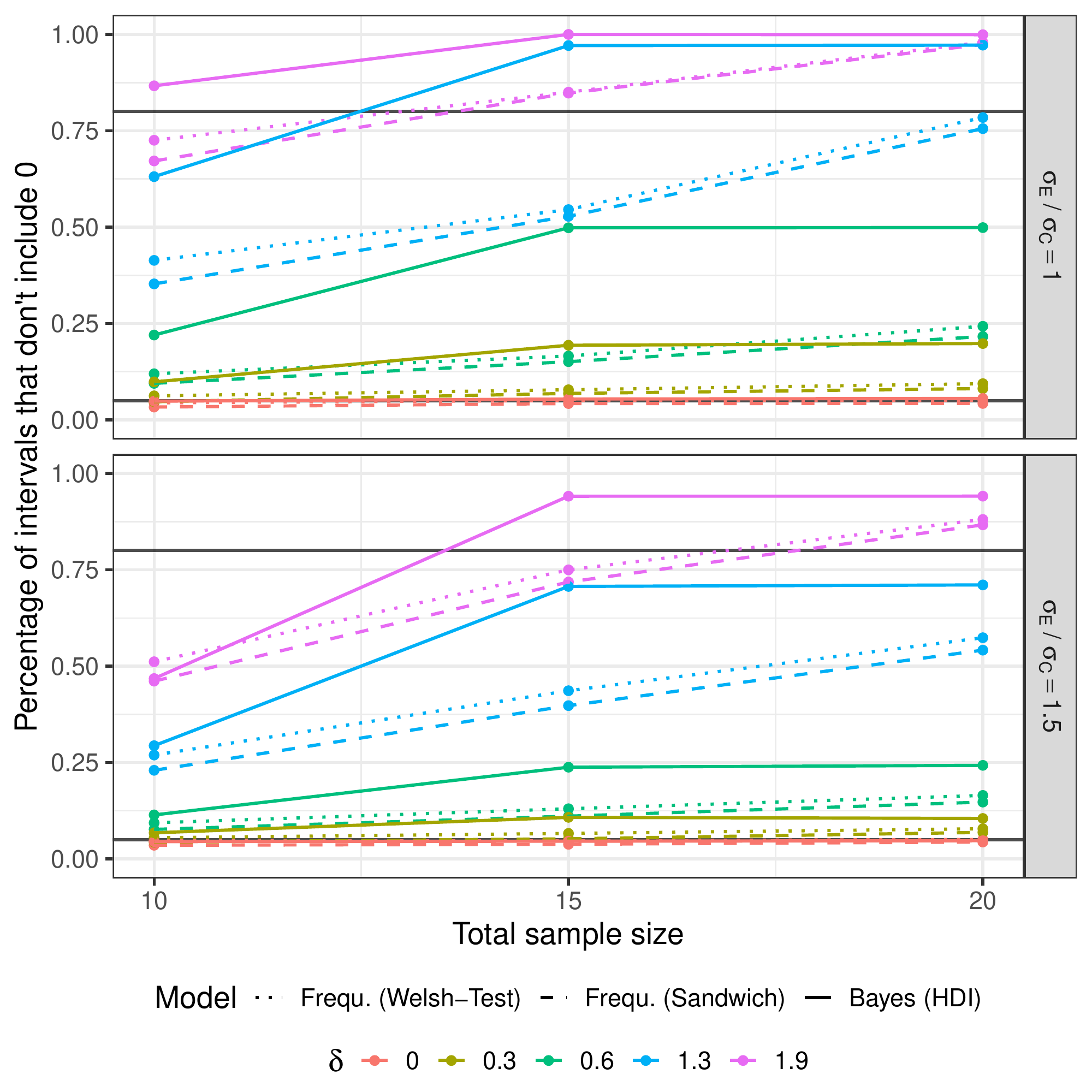}
		\caption{}
		\label{fig:Power_TypeI} 
	\end{subfigure}
	
	\begin{subfigure}[b]{0.55\textwidth}
		\includegraphics[width=1\linewidth]{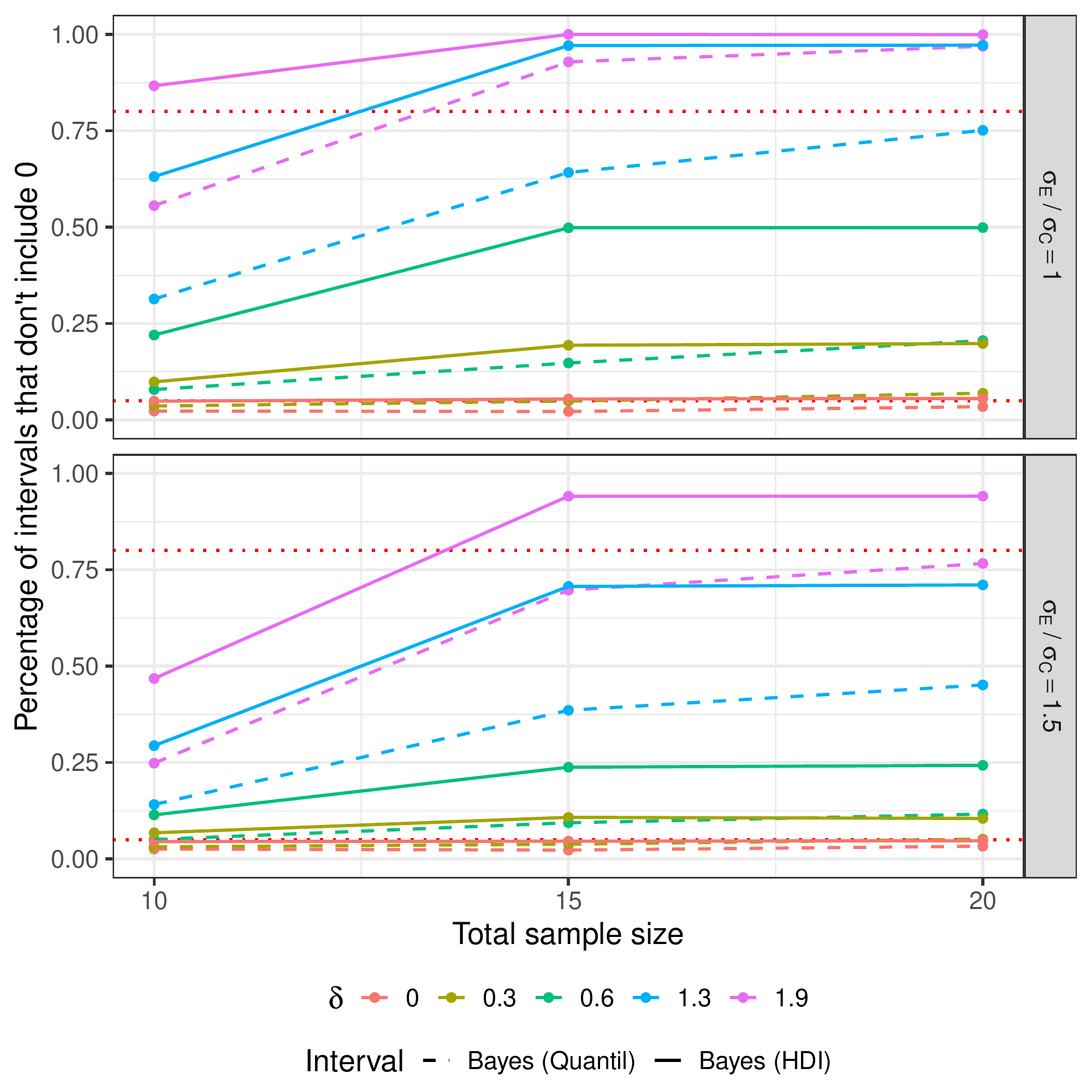}
		\caption{}
		\label{fig:Power_TypeI_Bayes}
	\end{subfigure}
	
	\caption{Proportion of the simulated data set in which the decision criteria for "success'' is met that the $95\%$ confidence or credible interval does not include the null value $0$ or that the p-value of the Welch test is smaller than $0.05$. The distributions are calculated over $10000$ simulated data sets. (a): Dotted line: proportion of simulated data sets with a 2-sided Welch test p-value$<0.05$. Dashed line: proportion of simulated data set where the frequentist confidence interval with the HC3 sandwich estimator doesn't include the value $0$. Solid line: proportion of simulated data set where the Bayesian highest density intervals (HDI) doesn't include the value $0$. (b): Comparison of the Bayesian quantile interval and HDI. Dotted lines: Conventional boundaries for the type I error rate or bower of 0.05 and 0.8.}
	\label{fig:OCmuCT}
\end{figure}

\begin{figure}[ht]
	\centering
	\includegraphics[height=0.6\textheight]{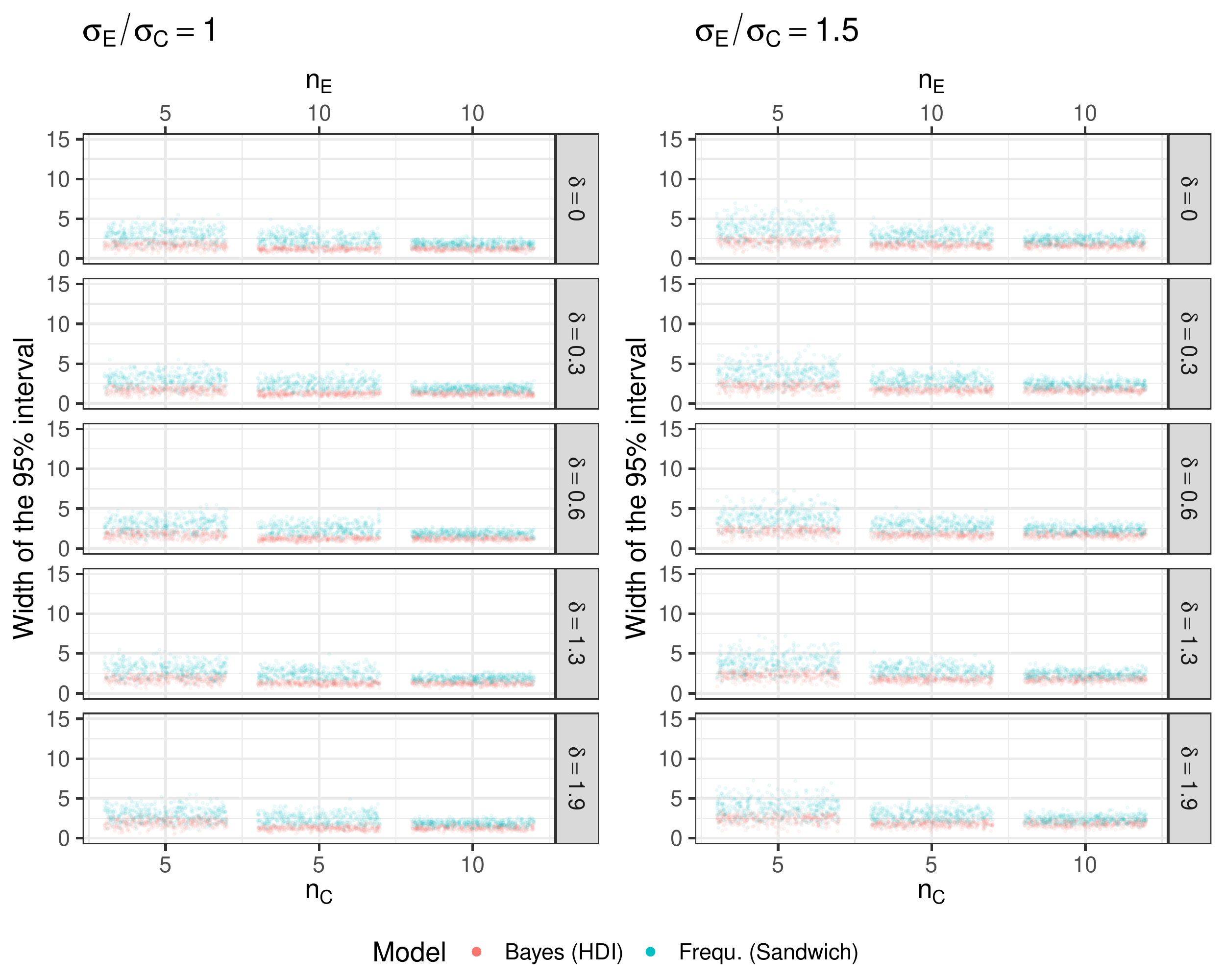}
	\caption{Widths of the $95\%$ Bayesian highest density intervals (HDI)s and the $95\%$ frequentist confidence intervals with the type HC3 sandwich estimator for a sub-sample of $500$ of the simulated data sets in different design setups.}
	\label{fig:CI_widths}
\end{figure}

\begin{figure}[ht]
	\centering
	\includegraphics[ height=0.6\textheight]{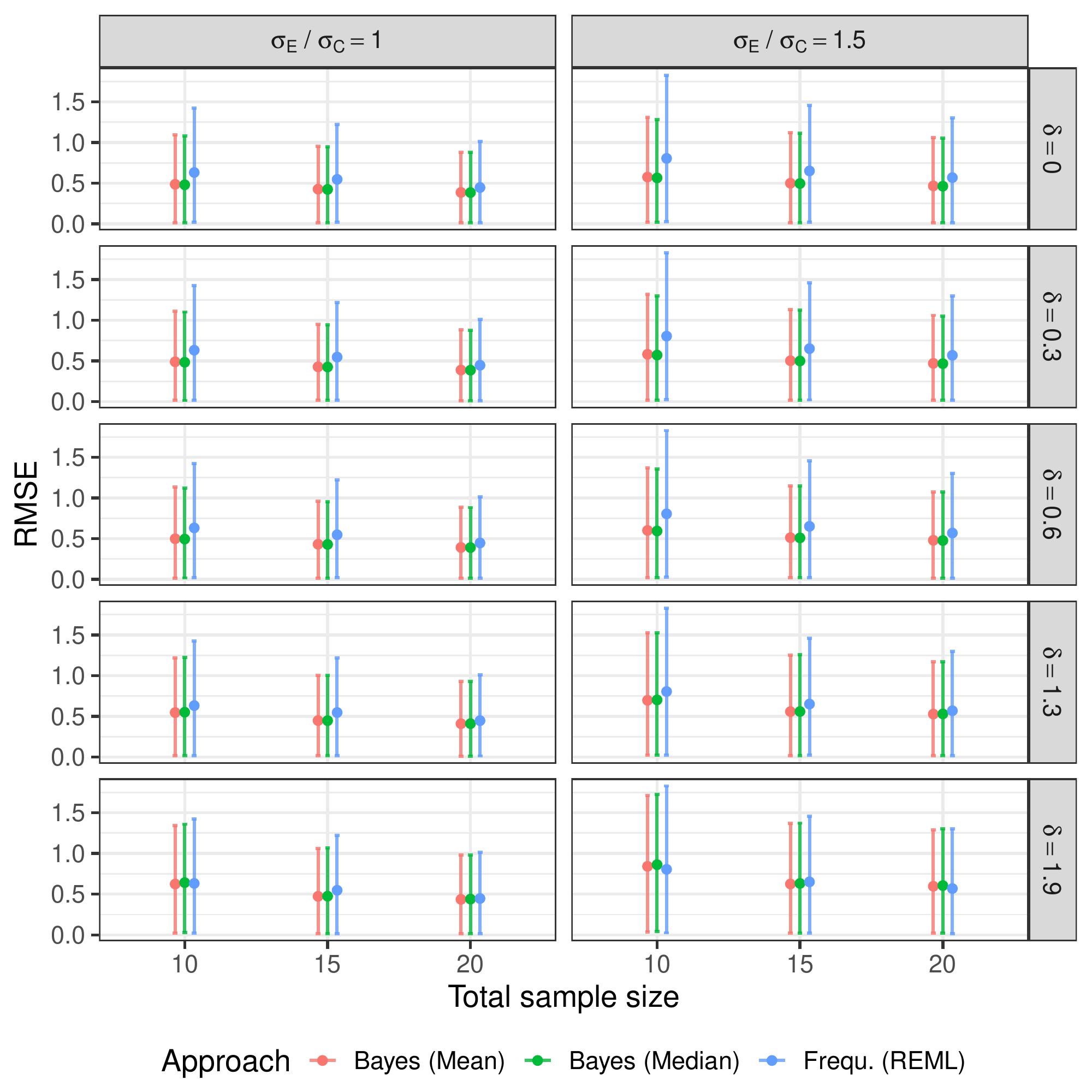}
	\caption{Root mean squared error (RMSE) calculated as root of the average squared difference of the point estimate of the treatment effect and the true mean of the treatment effect ($E(\delta)$). The RMSEs are represented as average over all simulated data sets together with $95\%$ quantile intervals.
 In the Bayesian model the point estimates of $\delta$ are arithmetic means and medians of the posterior MCMC draws of $\delta$ and in the frequentist model the estimates of $\delta$ are calculated as difference of the means in the treatment and control group.}
	\label{fig:rmse}
\end{figure}

\begin{figure}[ht]
	\centering
	\includegraphics[ height=0.6\textheight]{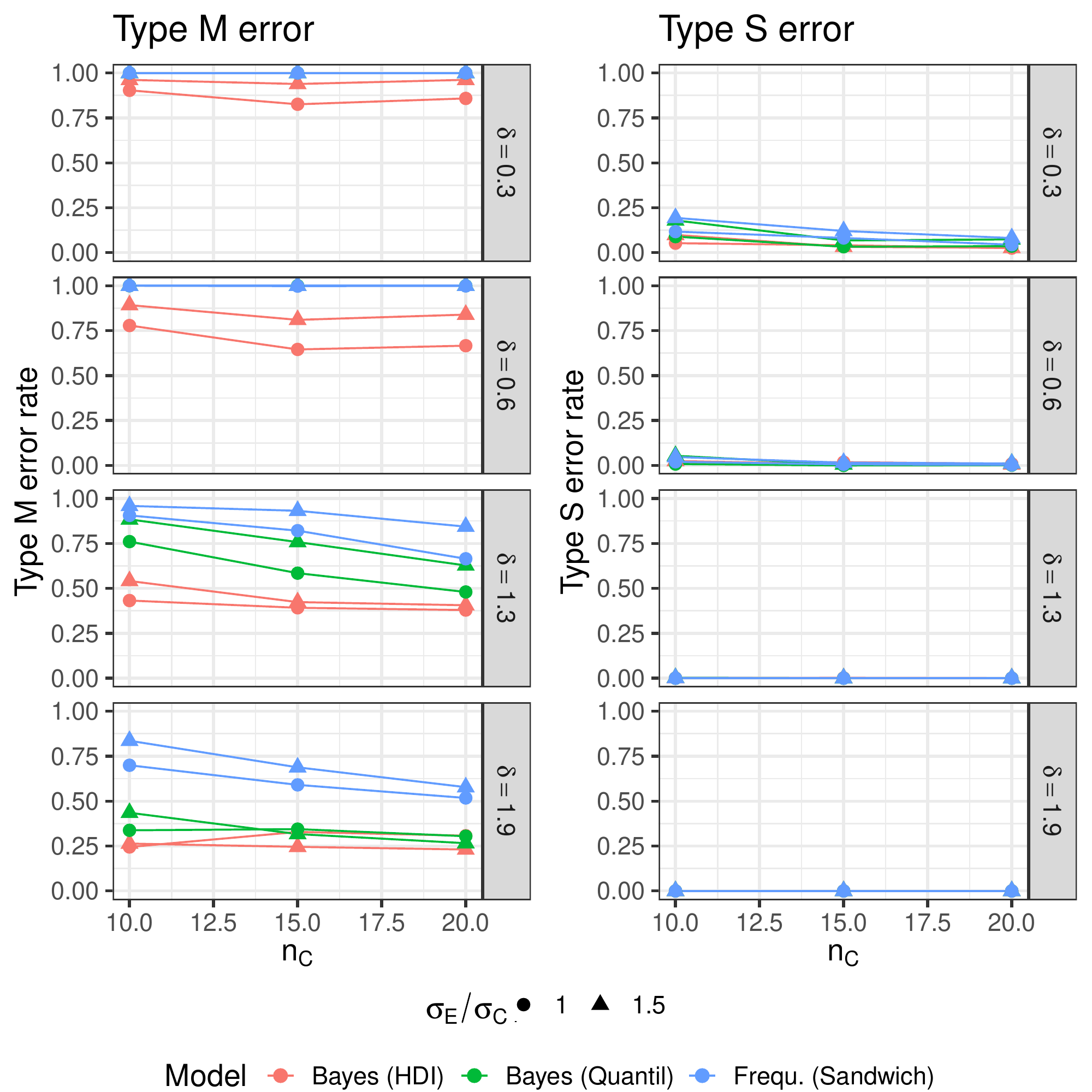}
	\caption{Type M (magnitude) error rate: Percentage of the simulated data sets where the effect estimate is bigger than the true treatment $\delta$ in absolute value, calculated in those data sets where the confidence/ credible interval did not include the value $0$.
		Type S (sign) error rate: Percentage of the simulated data sets where the effect estimate had a different sign than the true treatment $\delta$ in absolute value, calculated in those data sets where the confidence/ credible interval did not include the value $0$.
  The distributions are calculated over $10000$ simulated data sets.}
	\label{fig:TypM_TypS}
\end{figure}

\begin{figure}[ht]
	\centering
	\includegraphics[ height=0.6\textheight]{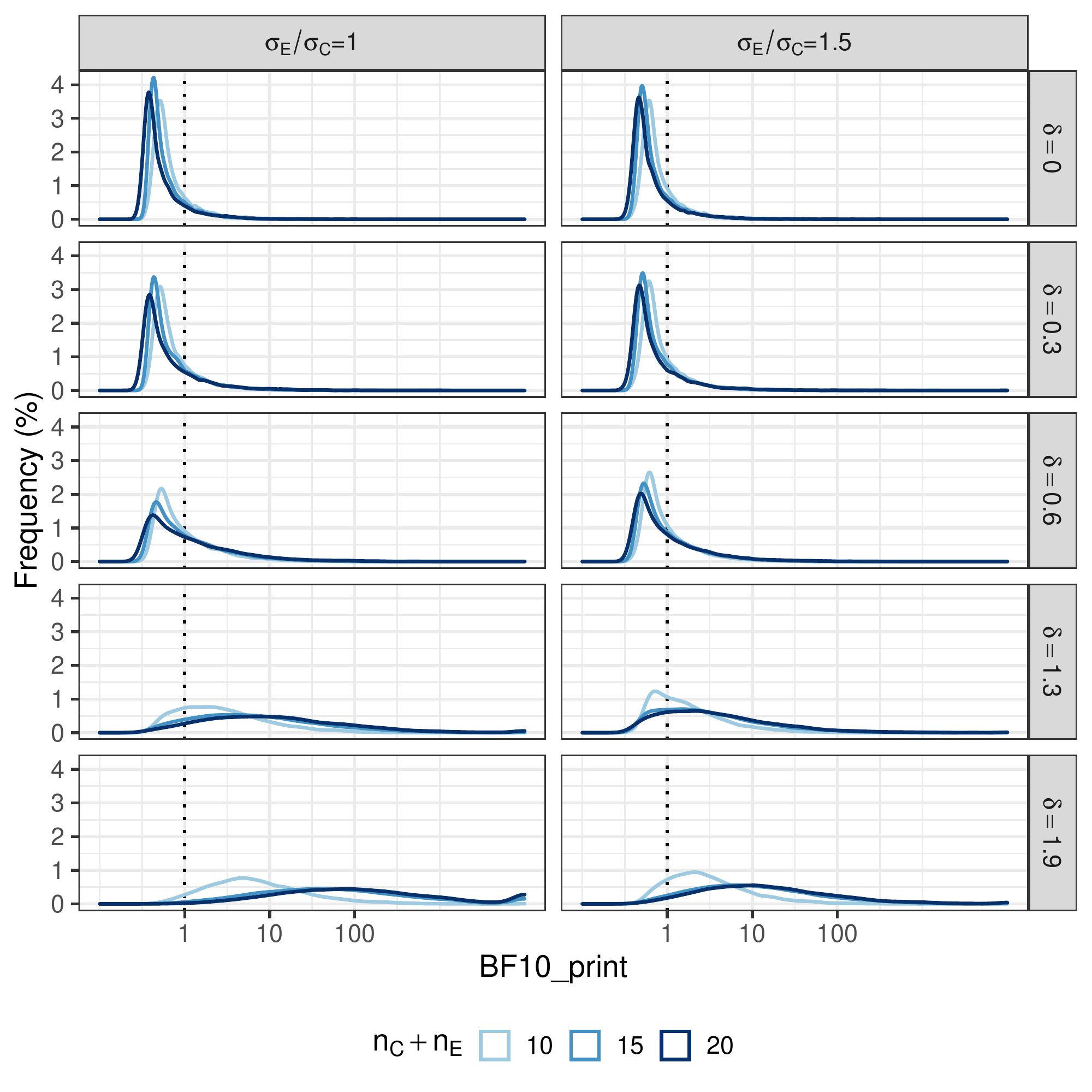}
	\caption{
 Distribution of the estimated Bayes factors $\text{BF}_{10}$ for evidence for the alternative model that the treatment effect $\delta$ is different from zero over the null model where it is equal to zero. 
 The distributions are calculated over $10000$ simulated data sets for different designs regarding the simulated true distribution in the data where for each design 10000 data sets are simulated.
 Extreme values of $\hat{\text{BF}}_{10}>10000$ are presented as $10000$ for representation.
	}
	\label{fig:BF10distribution_categories}
\end{figure}

$10000$ data sets are simulated with the chosen designs under model \eqref{eq:datsimbvtv}.
The results of the design analysis are presented in figures \ref{fig:OCmuCT}, \ref{fig:CI_widths}, \ref{fig:rmse}, \ref{fig:TypM_TypS} and \ref{fig:BF10distribution_categories}. The simulations were run on the High Performance Computing cluster of Baden-Wuerrtemberg (bwHPC). Using parallel computation, array jobs and the \texttt{update} functionality in \texttt{brms}, the computations took about 7 hours. 
In figure \ref{fig:OCmuCT} a) the number of p-values of the Welch test that are smaller than $0.5$, $95\%$ frequentist confidence intervals and $95\%$  Bayesian quantile credible intervals that don't include the null effect $\delta=0$ are shown, for the different experimental designs, as a function of the total sample size (number of animals in both the new experimental and control group, $n_E+n_C$). The results with the frequentist regression based decision and the p-values are rather similar with slightly higher rejection percentages with the Welch test.
Comparing the frequentist and the Bayesian curves in those designs that were simulated with unequal group means ($\theta_C\neq \theta_E$), the additional information in the prior distribution leads to a higher percentage of rejections for the Bayesian model than the frequentist model. Power in this context can be defined as percentage of $95\%$ confidence or rather credible interval that exclude the value null. 
In those designs, that were simulated with equal standard deviations ($\sigma_C=\sigma_E$), such a power of at least $80\%$ is reached with $n_E=10$ and $n_C=5$ or $n_C=10$ in those cases that were the effect was large with $\delta=1.9$ and also in the Bayesian model with $\delta=1.3$. In those designs, that were simulated with unequal standard deviations, a power of at least $80\%$ is only reached with $n_E=10$ and $n_C=10$ for $\delta=1.9$ and for $n_E=10$ and $n_C=5$ also for the Bayesian model.
Figure \ref{fig:OCmuCT} b) compares the curves of the Bayesian quantile intervals from a) to those derived from Bayesian highest posterior density intervals (HDI) (for details on HDI and quantile intervals see for example \cite{held2014applied}).

Figure \ref{fig:CI_widths} illustrates the precision for the designs as alternative goal for experimental planning. It shows the widths of a random sample of confidence or credible intervals as suggested by \cite{kruschke2015doing} and \cite{elsey2021powerful}. Using the type HC3 sandwich estimator to account for possibly different standard deviations in the experimental and the control group, many of the frequentist confidence intervals are much wider than the Bayesian ones from the distributional model. This can be observed especially for the smaller sample sizes with five animals in the control or experimental group and for actually unequal residual standard deviations $\frac{\sigma_E}{\sigma_C}=1.5$. In contrast, for ten animals in both groups and for actually equal residual standard deviations, the width of the frequentist intervals is more similar to that of the Bayesian intervals. 
If the designs are analyzed with regard to the goal to reach a certain precision instead of power, then a target precision has to be defined in terms of a threshold for the desired interval widths. For example, if the goal was that with a high probability (e.g. $95\%$) all intervals in the designs with equal standard deviations are not wider than a threshold of $1.1$, then this would be achieved for the larger sample sizes $n_E=10$ and $n_C=5$ or $n_C=10$ in the Bayesian model and for none of the sample sizes in the frequentist model.

Figure \ref{fig:rmse} shows that, for those designs with no to moderate effect $\delta$, the mean squared error (MSE) of the frequentist estimate is on average bigger than the MSE in the Bayesian model. For bigger sample sizes, the MSE in the frequentist model gets constantly smaller. In contrast, in the Bayesian model, the MSE only decreases slightly with sample size in the designs with the larger effect sizes.

Figure \ref{fig:TypM_TypS} shows the average type M and type S error rate for the different designs. In most designs, the type S error rates are quite similar in the Bayesian models compared to the frequentist models. Large differences exist however in the type M error rate of the designs with larger $\delta=1.3$ and $\delta=1.9$ where the type M error rate is notably larger in the frequentist models than in the Bayesian models. For the frequentist model, the type M error is very large in all designs, whereas for the Bayesian model it gets smaller with the larger effects since the choice of prior distribution results in posterior distributions for $\delta$ that are pulled towards zero. This indicates that, if the new experiment is conducted with a frequentist analysis and either of these designs (and if the new data is actually reflected by the model used for the fake data generation), then orienting the design choice and statements on statistical significance (in terms of whether or not the confidence or credible intervals didn't include the null value) almost always leads to an overestimation of the treatment effect. The type S error is small in all designs, but around $10\%$ with the models with the small effect $\delta=0.3$.

The analysis of the estimated Bayes factors gives an impression of how much evidence there is for the null and the alternative hypothesis.
The distributions of the estimated Bayes factors in the different designs are presented in figure \ref{fig:BF10distribution_categories}. In the designs with no effect ($\delta=0$) or small effect ($\delta=0.3$) the distribution of the Bayes factor has a clear peak and the majority of its probability mass below the value one, suggesting that based on the conservative choice of equal priors for the group means and the evidence from the data, $H_0$ is more likely than $H_1$. While for the smaller sample size ($n_C=n_E=5$) the peak of the distribution is closer to the value one, the peak moves further towards zero for bigger sample sizes since then there is more evidence for $H_0$ as compared to the case of smaller sample sizes. As compared to the null effect, the distribution of Bayes factors gets flatter for larger values of $\delta$ since then the information in the data starts to rule out the tendency of the prior evidence ratio to support the null hypothesis that states equality in the group means. For the very large effect $\delta=1.9$, the distribution of the Bayes factors has its peak above the value one, indicating that there is more evidence for $H_1$, while for the effects that are only of size $\delta=1.3$ the peak of the distribution is still very close to the value one, especially for the smaller sample sizes and the case of unequal standard deviations. For bigger sample sizes with $n_E=10$ the distribution of the Bayes factors becomes very flat with very extreme values.
 Table \ref{tab:mean_moderateh1} shows that in those designs, where the data was simulated under the null hypothesis with $\delta=0$, there is on average more evidence for $H_0$ and the posterior median model probability of $H_1$ is smaller than $50\%$. 
A goal for design analysis could be to find a sample size where the $95\&$ quantile interval of posterior model probability for $H_1$ does exceed the value $50\%$. This goal would be achieved in those designs with the very large effect of $\delta=1.9$ and with equal standard deviations in both groups, for sample sizes of ten animals in the experimental group and five or ten animals in the control group.
Another goal for design analysis could be to find a sample size that leads to a probability of the Bayes factor indicating at least moderate evidence for $H_1$ of at least $80\%$. For this goal a sample size of ten animals would be enough in those designs with $\delta=1.9$ that correspond to reversing the negative ovariectomy effect in the knockout animals on average.
 

\begin{table}[ht]
\centering
\begingroup\fontsize{9pt}{10pt}\selectfont
\begin{tabular}{rrrrllr}
  \hline
$\sigma_E / \sigma_C$ & $\delta$ & $n_C$ & $n_E$ & $p(\mathcal{M}_1|y)$ & $BF_\text{10}$ & $BF_\text{10}>3$ (\%) \\ 
  \hline
1.0 & 1.9 & 10.0 & 10 & 0.99 [0.73,1] & 92 [2.5,2.4e+15] & 96.6 \\ 
  1.0 & 1.9 & 5.0 & 10 & 0.98 [0.62,1] & 49 [1.6,1400000] & 93.6 \\ 
  1.5 & 1.9 & 10.0 & 10 & 0.92 [0.47,1] & 12 [0.86,1000] & 81.8 \\ 
  1.0 & 1.3 & 10.0 & 10 & 0.91 [0.36,1] & 9.6 [0.56,1400] & 74.1 \\ 
  1.5 & 1.9 & 5.0 & 10 & 0.9 [0.43,1] & 9.1 [0.75,580] & 76.1 \\ 
  1.0 & 1.9 & 5.0 & 5 & 0.86 [0.45,0.99] & 6 [0.82,150] & 72.4 \\ 
  1.0 & 1.3 & 5.0 & 10 & 0.85 [0.34,1] & 5.8 [0.51,420] & 64.9 \\ 
  1.5 & 1.3 & 10.0 & 10 & 0.74 [0.31,0.99] & 2.9 [0.46,170] & 49.3 \\ 
  1.5 & 1.9 & 5.0 & 5 & 0.71 [0.38,0.98] & 2.5 [0.61,43] & 43.1 \\ 
  1.5 & 1.3 & 5.0 & 10 & 0.71 [0.33,0.99] & 2.4 [0.49,98] & 43.9 \\ 
  1.0 & 1.3 & 5.0 & 5 & 0.7 [0.33,0.98] & 2.3 [0.49,57] & 41.7 \\ 
  1.5 & 1.3 & 5.0 & 5 & 0.58 [0.34,0.96] & 1.4 [0.51,23] & 24.0 \\ 
  1.0 & 0.6 & 10.0 & 10 & 0.45 [0.25,0.97] & 0.81 [0.32,31] & 18.4 \\ 
  1.0 & 0.6 & 5.0 & 10 & 0.43 [0.27,0.94] & 0.77 [0.38,16] & 14.5 \\ 
  1.5 & 0.6 & 5.0 & 5 & 0.43 [0.31,0.89] & 0.76 [0.45,7.8] & 8.1 \\ 
  1.0 & 0.6 & 5.0 & 5 & 0.42 [0.29,0.91] & 0.74 [0.4,9.8] & 10.4 \\ 
  1.5 & 0.6 & 5.0 & 10 & 0.42 [0.3,0.92] & 0.73 [0.42,11] & 10.8 \\ 
  1.5 & 0.6 & 10.0 & 10 & 0.42 [0.27,0.94] & 0.71 [0.37,17] & 12.5 \\ 
  1.5 & 0.3 & 5.0 & 5 & 0.41 [0.3,0.84] & 0.69 [0.44,5.1] & 5.0 \\ 
  1.5 & 0.0 & 5.0 & 5 & 0.4 [0.3,0.81] & 0.67 [0.44,4.3] & 3.8 \\ 
  1.0 & 0.3 & 5.0 & 5 & 0.38 [0.28,0.83] & 0.6 [0.39,4.7] & 4.7 \\ 
  1.5 & 0.3 & 5.0 & 10 & 0.37 [0.29,0.83] & 0.6 [0.41,4.8] & 4.6 \\ 
  1.0 & 0.0 & 5.0 & 5 & 0.36 [0.28,0.76] & 0.57 [0.38,3.2] & 2.8 \\ 
  1.5 & 0.0 & 5.0 & 10 & 0.36 [0.29,0.76] & 0.57 [0.41,3.3] & 2.8 \\ 
  1.5 & 0.3 & 10.0 & 10 & 0.36 [0.26,0.87] & 0.55 [0.36,6.5] & 5.5 \\ 
  1.0 & 0.3 & 5.0 & 10 & 0.35 [0.27,0.83] & 0.53 [0.36,4.9] & 4.6 \\ 
  1.5 & 0.0 & 10.0 & 10 & 0.34 [0.26,0.79] & 0.53 [0.36,3.8] & 3.3 \\ 
  1.0 & 0.0 & 5.0 & 10 & 0.33 [0.26,0.72] & 0.49 [0.36,2.6] & 2.0 \\ 
  1.0 & 0.3 & 10.0 & 10 & 0.33 [0.23,0.88] & 0.48 [0.31,7.3] & 6.0 \\ 
  1.0 & 0.0 & 10.0 & 10 & 0.3 [0.23,0.76] & 0.43 [0.3,3.1] & 2.6 \\ 
   \hline
\end{tabular}
\endgroup
\caption{
Alternative quantification of evidence for the model under $H_1$ in the different designs (as represented by the four columns to the left). $p(\mathcal{M}_1|y)$: posterior probability for model $\mathcal{M}_1$ as model under $H_1$.
$\text{BF}_{10}$: Median Bayes factor $\text{BF}_{10}$ with 95\% quantile interval. $BF_\text{10}>3$: percentage with at least moderate evidence for $H_1$ as categorized by \cite{jeffreys1998theory}. The distributions are calculated over $10000$ simulated data sets-} 
\label{tab:mean_moderateh1}
\end{table}

Further designs were evaluated that are not represented here. More specifically, the effect of decreasing the standard deviation in the prior for $\theta_C$ to the half of its previous size was examined with the aim to make it more informative. However, this did not have a noticeable effect. Additionally, the effect of increasing the standard deviation for the prior of $\delta$ to two times its previous value and twenty times its previous value was examined, with the aim to make it less informative. This lead to a higher FDR since the prior had less effect on the posterior and extreme observations in the small data set could lead to false positive claims. Furthermore, the prior distribution with a standard deviation of twenty times its previous value ($0.7\cdot20=14$) lead to a very flat distribution of Bayes factors even for those designs that were simulated with a truly large or very large effect size. Of note, non-informative priors are not recommended to be used in the context with Bayes factors \cite{schad2022workflow}.
Moreover, fake-data was simulated under a Bayesian design with pobability distributions for all parameter. There, the curves of the designs that where simulated with a null effect on average ($E(\delta)=0$) showed, that, if the effect has a large standard deviation (like $SD(\delta)=1.2$), the percentage of intervals that doesn't include the null effect gets much larger than $5\%$. Hence, one would make many more type I error as usually intended in the frequentist framework in the cases with total sample sizes $15$ and $20$ and in groups that have on average the same relative bone volume but with the experimental group having larger standard deviations. Since in the Bayesian simulation framework neither the frequentist nor the Bayesian interval based decision rule was designed for having (asymptotically) such a type I error rate smaller than $5\%$, this error rate may get much larger than $5\%$.


For comparison, classical sample size calculation by solving power equalities is carried out with the frequentist Welch test and the power\_t\_test() function in the R \texttt{MESS} package (\cite{r2020mess}). As candidates for the effect sizes, for the group-specific standard deviations and for allocation ratio to the control and experimental group ($\frac{n_E}{n_C}$) the same design settings as in table \ref{tab:Designs_muCT} are examined. The significance level and target power are set to the conventional values of 0.05 and 0.8.
The results are presented in table \ref{tab:ssfreq} According to this calculation, a sample size of six animals in both experimental and control group would be sufficient to detect an effect (as difference in the means) of at least the size $1.9$ with a power of at least $80\%$ in this test, if the residual standard deviations in both groups are equal ($\sigma_E / \sigma_C=1$) and the allocation ratio to both groups is also equal ($n_E /n_C=1$) (setting 13). If about twice the animals shall be assigned to the treatment group, then $n_C=5$ animals for the control group and $n_E=9$ animals for the experimental group are calculated for detecting at least this effect size and equal standard deviations (setting 14). For the same $\delta_\text{rel}=1.9$, but a greater standard deviation in the experimental group ($\sigma_E=1.5$), nine animals are calculated for both groups for an equal allocation ratio (setting 15) and six and eleven animals for an allocation ratio of twice the amount to the experimental group (setting 16).

\begin{table}[ht]
\centering
\begin{tabular}{rrrrrrr}
  \hline
Setting & $\delta_\text{rel}$ & $\sigma_C$ & $\sigma_E$ & Alloc. ($\frac{n_E}{n_C}$) & $n_C$ & $n_E$ \\ 
  \hline
1 & 0.3 & 1.0 & 1.0 & 1 & 176 & 176 \\ 
  2 & 0.3 & 1.0 & 1.0 & 2 & 132 & 264 \\ 
  3 & 0.3 & 1.0 & 1.5 & 1 & 285 & 285 \\ 
  4 & 0.3 & 1.0 & 1.5 & 2 & 187 & 373 \\ 
  5 & 0.6 & 1.0 & 1.0 & 1 & 45 & 45 \\ 
  6 & 0.6 & 1.0 & 1.0 & 2 & 34 & 68 \\ 
  7 & 0.6 & 1.0 & 1.5 & 1 & 72 & 72 \\ 
  8 & 0.6 & 1.0 & 1.5 & 2 & 48 & 95 \\ 
  9 & 1.3 & 1.0 & 1.0 & 1 & 11 & 11 \\ 
  10 & 1.3 & 1.0 & 1.0 & 2 & 9 & 17 \\ 
  11 & 1.3 & 1.0 & 1.5 & 1 & 17 & 17 \\ 
  12 & 1.3 & 1.0 & 1.5 & 2 & 11 & 22 \\ 
  13 & 1.9 & 1.0 & 1.0 & 1 & 6 & 6 \\ 
  14 & 1.9 & 1.0 & 1.0 & 2 & 5 & 9 \\ 
  15 & 1.9 & 1.0 & 1.5 & 1 & 9 & 9 \\ 
  16 & 1.9 & 1.0 & 1.5 & 2 & 6 & 11 \\ 
   \hline
\end{tabular}
\caption{Results from a classical sample size calculation with a two-sided Welsh test, a power of 0.8 and a significance level of 0.05 calculated with the power\_t\_test() function in the R MESS package. $\delta_\text{rel}$: Minimal clinically relevant effect size that shall be detected by the test.  $\sigma_C$, $\sigma_E$: (estimated) standard deviations in the new experiment. Alloc. ($\frac{n_E}{n_C}$): allocation ratio for the mice in the new experiment to the experimental and control group. $n_C$, $n_E$: resulting sample sizes for the new experiment.} 
\label{tab:ssfreq}
\end{table}

	\section{Discussion}

 \subsection{Summary}

	In this work aspects of sample size determination and analysis of translational animal experiments in a Bayesian framework were discussed and compared to the classical frequentist procedure in a null hypothesis significance testing (NHST) framework. The considerations where illustrated on a real-world animal experiment examining the knockout effect of the C5aR1 receptor in osteoclasts and osteoblasts on the relative bone volume (C5aR1 example). The determination of a sample size depends on the model assumptions on the new experiment and on the statistical goal of the analysis and model fit. In the Bayesian framework these assumptions include prior distributions for the model parameters. As basis for setting up the prior distributions, a Bayesian meta-analysis model was estimated to available historical data, consisting of internal data from the applicant of the new animal experiment and from external data from the Mouse Phenome Database (MPD). For comparison, also a frequentist model was fit that gave quite similar point estimates.
Design analysis was performed with prior distributions and fake-data that was based on the fitted meta-analysis model. The estimate of the effective sample size of the meta-analytic predictive prior for the control group in the new experiment indicated that the historical control data was only worth two animals, which corresponds to the general impression that sample size planning in translational animal experiments often comes with large uncertainties (see \cite{mayer2013limitierte}). As sample size candidates for the design analysis, the minimum and maximum of the range of typical sample sizes for preclinical translational animal experiments were considered. The range of the candidates for the treatment effect was chosen based on what seemed realistic according to the knowledge from the meta-analysis model that was fitted to the historical data. The simulations required large computational resources, especially when Bayes factors are evaluated (here the High Performance Computing cluster of Baden-Wuerrtemberg (bwHPC) was used and computations in the 30 designs with each 10000 simulated data sets took about 7 hours using parallel computation, array jobs and the \texttt{update} functionality in the \texttt{brms} package). In this example the power-based sample size calculation (here done with a Welch test) suggested that eleven or less animals in both groups would be enough to detect differences in the means of at least $1.3$ for an equal allocation ratio of the animals to both groups and residual standard deviations of one in both groups and nine or less for differences of size $1.9$ or greater or rather six and eleven animals for an unequal allocation ration and larger standard deviations in the experimental group. 
 However, the analysis the type M error rates showed that the design and analysis with classical frequentist can lead to a high percentage of overestimated effect sizes in those cases where the analysis of the data in the new experiment results in a test decision against the null hypothesis. 
Using as Bayes factors oriented goal that $95\%$ of the posterior probability of the model under $H_1$ is above the value $50\%$ (representing equal model probability for both the model under $H_0$ and $H_1$), only in those designs with equal standard deviations of one and an effect of size $1.9$, ten animals in the experimental group (and five animals or then in the control group) would be enough. Using goal that the Bayes factor exceeds with $80\%$ probability the heuristic threshold value for moderate evidence for $H_1$ defined by \cite{jeffreys1998theory,lee2014bayesian}, then a sample size of ten animals in both groups would also be enough for unequal standard deviations with a standard deviation in the experimental group of $1.5$ in the new experiment and a standard deviation in the control group of $1$.

 	Several chances and challenges were identified in this Bayesian meta-analytic predictive framework as compared to the classical frequentist framework. 
 If the goal of planning the new experiment is to achieve a certain statistical power, the use of the historical data did not lead to a lower sample size in a Bayesian analysis with prior predictive distributions from the historical data than 
 with classical frequentist power calculations. However, the use of fake-data design analysis based on the historical data and the evaluation of additional model characteristics and statistical allowed a better representation and estimation of present uncertainties in the model parameters.
	More specifically, the Bayesian model framework allows to formally incorporate a priori knowledge (deduced from historical data) as prior distribution in the analysis of a new experiment. Secondly, uncertainties, that are almost always present in research stage as early as preclinical translational research, are better represented by modeling all of the model parameters as random variables instead of fixed parameters. Thirdly, the estimation with MCMC methods allows also for more complex models, that might represent some data more accurately. As an example, different residual standard deviations were modeled for different experimental groups in the historic and the new data of the C5aR1 example and the estimates seemed more precise than frequentist sandwich estimators. Fitting a meta-analysis model to the historic data provides a quantitative summary and can be used to define the prior distributions for the new experiment. With Bayesian methods heterogeneity can be reflected in the means of different historic experiments, also in the case of only few experiments or groups. In contrast, frequentist models can deal less well with fitting meta-analysis or hierarchical models in heterogeneous data in the situation of with few, small experiments \cite{gelman2006prior, friede2017meta2studies,friede2017metaFewSmall}.
Meta-analysis of similar historical experiments not only provides an initial guess, that can be used for prior specification, but also a tool for quality control. More specifically, flaws in the experimental design or analysis may stick out or statements may have to be relativized when some measurements originating from a common mean hierarchical model differ significantly from supposedly related measures \cite{walley2016using}.
 Planning and analyzing the new experiment's in the bigger of the estimated meta-analysis model of the historical data may help to make more appropriate statements and may lead to more reproducible results as a step out of the reproducibility crisis in animal research \cite{ioannidis2005most, begley2012raise, begley2015reproducibility, loken2017measurement, goodman2016does, jilka2016road, freedman2015economics, macleod2019reproducibility, voelkl2020reproducibility}.
   The consideration of not only internal but also external data like from the MPD gives a broader picture of the natural variation of the outcome of interest. This helps to make more generalizable statements that are potentially more likely to being translated to the application in humans or to the reproduction of experiment results in other animals as suggested by \cite{voelkl2016reproducibility,voelkl2020reproducibility}.
   The point estimates of the Bayesian and frequentist meta-analysis model in this application example were quite similar, but the Bayesian approach allowed an easier estimation of the confidence intervals as quantiles of the posterior draws. The frequentist and Bayesian estimates might differ more if heterogeneity was model for a grouping variable with smaller number of groups. This could be the case if also the data laboratory (MPD, internal) was modelled. In this case the Bayesian approach has proven to be superior \cite{friede2017metaFewSmall,friede2017meta2studies}.
   Furthermore, determine sample size by design analysis using fake-data simulation instead of the classical determination by power inequalities, addresses several problems that are common in translational research. In particular, the variation in an experimental design and data may be better represented by a continuous value as the Bayes factor instead of the outcome of a binary decision rule and hence it may be of more value to just report the Bayes factors associated with the different designs. In particular, \cite{schad2022workflow} show in the context of a standard cognitive experiment that many standard designs don't have sufficient evidence for making conclusive decisions and support the idea of increasing the sample sizes by sharing data across different researchers and laboratories.

	Concerning the challenges, fitting Bayesian models with MCMC methods requires at least a basic understanding of the additional convergence diagnostics to ensure a proper model fit. Other recommended steps are prior and posterior predictive checks to make sure that the specified prior and posterior distributions are reasonable. The steps and decisions that are commonly made in a Bayesian framework are described as a Bayesian workflow \cite{gelman2020bayesian} and may become quite complex. In particular, Bayesian inference has typically many more determining factors than frequentist inference through the specification of all model parameters' prior distributions and MCMC parameters. The problem is even worse when Bayesian methods are considered in an design analysis framework where additional determining factors come with different design options. This abundance of determining factors makes it hard to understand the effect of changing single determining factors for the posterior inference and makes the investigation of all combinations of determining factors becomes soon incomprehensible. Furthermore, there is so far no consensus in literature about which procedure to use for sample size determination in Bayesian framework (if and what decision functions and thresholds should be used etc.). In the case where only few, small previous experiments are available, special attention has to be paid to the assumptions on the prior model and its hyperparameters. Especially, setting up a reasonable model for the heterogeneity parameter is important to properly reflect the variation in the background population under focus and prevents from driving overly-confident claims that only apply to standardized animals in a single experiment.
Concerning the use of Bayes factors for design analysis and sample size determination, challenges are firstly the definition of a threshold. 
 Secondly, Bayes factors are highly sensitive to the choice of prior distributions as shown by \cite{schad2021toward} and the usual estimation method by bridge sampling or the Savage-Dickey method requires a large number of MCMC iterations to be stable \cite{gronau2020bridgesampling} and preferably several repeated estimations. This makes the estimation of Bayes factors also computationally challenging. Finally, possible bias in the Bayes factors estimate should be examined in simulation based calibrations (SBC) \cite{schad2021toward}. These Bayes factor workflow procedures again increase the already high manual and computational burden of the simulation based design analysis which might also constitute an obstacle for the application in translational research where the resources of the researchers are often quite limited.
 With regard to meta-analysis, challenges are that it is difficult to find the relevant historic information since the rate of annual publications in preclinical research is very high \cite{bannach2021technological} and the published estimates are often subject to bias like publication bias \cite{sena2010publication,ter2012publication,conradi2017publication,bih2021EMBARC}.
 Furthermore, the relevant literature is often unorganized and outcomes do not follow a unique terminology what makes it hard to compare results from different experiments \cite{smith2005mammalian}. 
 These facts make it currently challenging for researchers of translational animal experiments to understand and correctly apply Bayesian methods and make sample size determination too extensive for practical applications in this context without the development of routines and applications that facilitate their use.

\subsection{Extensions}
There are several extensions to the here presented methodology for planning and analysing translational animal experiments using Bayesian meta-analytic predictive approaches and fake-data design analysis.
  In this work the distribution of the Bayes factors were used to visualize the evidence ratio for the null and alternative hypothesis under different models. To transform the distribution into a decision function, a heuristic threshold was defined based on a classification scheme of Jeffreys \cite{jeffreys1998theory} or by checking if the $95\%$ posterior model probability for the model under $H_1$ ($p(\mathcal{M}_1|y)$) did exceed the value $50\%$ (representing equal probability for model the model undeer $H_1$ and under $H_0$. A more systematic approach to setting a threshold for the Bayes factors is by the definition of utility functions as illustrated by
  \cite{schad2021toward} and \cite{schonbrodt2018bayes}. 
 Bayes factors compare the ``out-of-sample" predictive performance of the two contrasting models (here the model under $H_0$ and under $H_1$). A further approach to making a decision whether or not there is evidence in the data that the effect $\delta$ differs from that stated by the null hypothesis is to compare the out-of-sample predictive performance by the investigation of posterior predictions.
  A common utility function that measures the out-of-sample predictive performance of a model is the expected log pointwise predictive density (ELPD) (for details see \cite{gelman2014understanding} and for practical estimation in a Bayesian framework see \cite{vehtari2017practical}).

  In this work a Bayesian meta-analysis model was fit to the historical data with the purpose to get prior distributions and to define a reasonable design analysis setups. 
Instead of the Bayesian meta-analysis model, also the estimates from a frequentist meta-analysis model can be used to set up parametric distributions for definition prior distributions and sampling distributions for the fake-data design analysis. This was done for example by \cite{schad2022workflow} in the context of simulations for the examination of the behavior of Bayes factors under different hypothesis.
In this work age and the historical data's experiment/ laboratory effects could not be incorporated since necessary data was missing. 
If only few historical data is available, it is difficult to check assumptions corresponding to a normal distribution and non-parametric models bay be more appropriate \cite{konietschke2021small}. Burr and Doss \cite{burr2005bayesian} suggest a Bayesian semi-parametric meta-analysis model that models the experiment-specific effects through a version of the Dirichlet process prior and implement it in the \texttt{R} package \texttt{bspmma} \cite{bspmma2012}. This model could be used if the normal assumption is in doubt or it can be compared to a parametric model using empirical Bayes to decide which model seems more appropriate \cite{bspmma2012}.
  There are general efforts for facilitating the use of systematic reviews and meta-analysis in animal research. Examples of such efforts include CAMARADES (Collaborative Approach to Meta-Analysis and Review of Animal Data from Experimental Studies) \cite{CAMARADES2021} which offers methodological advise and tools for conducting systematic reviews and meta-analysis in animal trials or the online review platform SyRF (Systematic Review Facility) \cite{bahor2021development}. Moreover, there are efforts for collecting animal data in common big databases \cite{eppig2005mouse, blake2006mouse,consortium2007Integration, hancock2008casimir, rita, pognan2021etransafe}. 
   Furthermore, there are advancements towards a mandatory (pre)registration for animal trials which could increase the amount and quality of public available data, reduce bias \cite{chamuleau2018translational,bert2019refining, heinl2019rethinking,baker2019animal,van2020publication}. With these advancements it is realistic, that in future better meta-analysis models of historical evidence can be fit. 
The simulated fake-data in the design analysis represent assumptions on the data in the new experiments. The assumptions are based on historical data that is summarized in a meta-analysis model. In the ideal case of prior distributions that match the true distribution of the future data, the prior represents just additional information that makes the posterior estimates more precise. If, however, the prior predictive distribution of the data places major parts of its probability mass to regions that are unlikely according to the empirical distribution of the observed data, there is so-called \emph{prior-data conflict}. The effects of prior-data conflict can be a problem when they invalidate the inference based on the posterior draws (see for example \cite{box1980sampling,evans2006checking}). To examine the effects of prior-data conflict for the chosen prior distributions on the posterior inference, data that deviates from the prior predictive distribution can be simulated. 
Further steps are necessary to get a better understanding and better visualizations of all determining factors that affect the posterior inference in the design analysis with Bayesian models.
Recently, the R package \texttt{priorsense} \cite{kallioinen2021detecting} has been developed which intends to help investigate the impact of the prior with respect to observed data. Packages like this might help to get a better feeling of how the chosen prior distributions affect the necessary sample size and test decisions.
If prior-data conflict seems to be a problem with the chosen prior distributions, the prior distributions can for example be robustified by adding less informative mixture components to the respective distributions, as suggested by \cite{schmidli2014robust}.

Instead of using data-based priors for the design and analysis of the the new experiment, the priors can also be chosen by prior elicitation \cite{garthwaite2005statistical,o2006uncertain}. In a prior elicitation approach, the analyst does not specify the prior directly (like here based on historical data). Instead, there is an subject expert that describes properties of the outcome of interest and the task of the analyst is to formalize this description as a prior distribution. Prior elicitation could be an appealing alternative for the here used data-based priors, especially When there is no useful historical data available. However, at the current state, technical, practical and societal challenges hinder the use of prior elicitation \cite{mikkola2021}.

	\section*{Acknowledgments}
 The author acknowledges support by Melanie Haffner-Luntzer from the Institute of Orthopedic Research and Biomechanics in Ulm, Germany, for providing animal data and for discussing the use of animal phenotype databases.
	Furthermore, the author acknowledges support by the state of Baden-Württemberg through bwHPC.
 
		\section*{Conflicts of interest}
	All author declares that they have no potential conflicts of interest.
	
	\section*{Funding}
	This work was funded by the German Federal Ministry of Education and Research (BMBF), grant number 031L0233.

\bibliographystyle{agsm} 
\bibliography{quellen}  






\end{document}